\documentclass[reprint,
superscriptaddress,
nobibnotes,
amsmath,amssymb,
aps,
]{revtex4-2}

\usepackage[utf8]{inputenc} 
\usepackage[T1]{fontenc}    
\usepackage{hyperref}       
\usepackage{url}            
\usepackage{amsfonts}       
\usepackage{amsmath,amsthm,amssymb} 
\usepackage{bm}
\usepackage{bbm}
\usepackage{dcolumn}
\usepackage{nicefrac}       
\usepackage{gensymb}
\usepackage{graphicx}
\usepackage{xcolor, etoolbox}
\usepackage{mathtools}
\usepackage{mathrsfs}
\usepackage{ragged2e}
\usepackage{microtype}      
\usepackage{braket}      
\usepackage{enumitem}
\usepackage{xspace}
\usepackage{dsfont}
\usepackage{float}

\newcommand{\angstrom}{\text{\normalfont\AA}}

\DeclarePairedDelimiter\abs{\lvert}{\rvert}
\renewcommand{\vec}[1]{\boldsymbol{#1}}
\newcommand\varpm{\mathbin{\vcenter{\hbox{%
  \oalign{\hfil$\scriptstyle+$\hfil\cr
          \noalign{\kern-.3ex}
          $\scriptscriptstyle({-})$\cr}%
}}}}
\newcommand\varmp{\mathbin{\vcenter{\hbox{%
  \oalign{$\scriptstyle({+})$\cr
          \noalign{\kern-.3ex}
          \hfil$\scriptscriptstyle-$\hfil\cr}%
}}}}

\hypersetup{
    colorlinks,
    linkcolor=blue,
    citecolor=blue,
    urlcolor=blue,
    breaklinks=true 
}

\graphicspath{{./figures/}} 

\begin{document}
\title{Analog of cosmological particle production in Dirac materials}

\author{Mireia Tolosa-Sime\'on}
\affiliation{Institut f\"ur Theoretische Physik III, Ruhr-Universit\"at Bochum, D-44801 Bochum, Germany}

\author{Michael M. Scherer}
\affiliation{Institut f\"ur Theoretische Physik III, Ruhr-Universit\"at Bochum, D-44801 Bochum, Germany}

\author{Stefan Floerchinger}
\affiliation{Theoretisch-Physikalisches Institut, Friedrich-Schiller-Universit\"at Jena, Max-Wien-Platz 1, D-07743 Jena, Germany}

\begin{abstract}
Two-dimensional van der Waals materials have recently been established experimentally as a highly-tunable condensed matter platform, facilitating the controlled manipulation of band structures and interactions.
In several of these materials, Dirac cones are present in the low-energy regime near the Fermi level. 
Thus, fermionic excitations emerging in these materials close to the Dirac cones have a linear dispersion relation near the Fermi surface as massless relativistic Dirac fermions. 
Here, we study low-energy fermionic excitations of such Dirac materials in the presence of a mass gap that may be generated by symmetry breaking. 
Introducing a dynamical Fermi velocity and/or time-dependent mass gap for the Dirac quasiparticles, we exhibit the emergence of an analog of cosmological fermion pair production in terms of observables such as the expected occupation number or two-point correlation functions. 
We find that it is necessary and sufficient for quasiparticle production that only the ratio between the mass gap and the Fermi velocity is time-dependent.
In this way, we establish that highly-tunable Dirac materials can serve as analog models for cosmological spacetime geometries, in particular, for Friedmann-Lema\^itre-Robertson-Walker expanding cosmologies.
We briefly discuss possibilities for experimental realization.
\end{abstract}

\maketitle

\section{Introduction}
\label{sec:Introduction}

Condensed matter systems can be used in various scenarios to emulate and study phenomena from completely different fields of physics such as elementary particle physics or gravity \cite{Volovik2010,Barcelo2011}. 
Such analog condensed matter models provide a novel perspective to approach questions that are not directly accessible in the original systems as they can potentially be realized experimentally in a well-controlled setup.
An important example for successful condensed matter realizations of analog gravity scenarios are Bose-Einstein condensates realizing expanding spacetime geometries by exploiting the highly tunable environment provided by cold-atom setups~\cite{Novello2002,Barcelo2003,Barcelo2003b,Fischer2004,Fischer2004b,Uhlmann2005,Calzetta2005,Weinfurtner2009,BECpaper,Viermann2022}.
Similar studies for the case of fermions are still missing.

Recently, two-dimensional moir\'e materials, such as twisted bilayer graphene (TBG), have been established as another highly tunable condensed matter platform allowing us to manipulate electronic band structures and interaction effects in a controlled manner~\cite{MacDonald2011,LopesDosSantos2012,Cao2018,Cao2018c,kennes2021moire,zhang2023}.
A key feature of moir\'e materials is that electron bands near the Fermi level can be tuned to become very narrow. 
This leads to an enhancement of interaction effects and tentatively supports the formation of correlated states, which have been confirmed experimentally, see, e.g., Refs.~\cite{Cao2018,Cao2018c,kennes2021moire,zhang2023,Yankowitz2019,Choi2019}.
Some moir\'e materials, including TBG but also $\Gamma$-valley transition metal dichalcogenide bilayers~\cite{Angeli_2021}, even belong to the class of Dirac materials~\cite{Wehling2014,Vafek2014}, i.e. they are characterized by the presence of fermionic low-energy excitations, described by a quasirelativistic Dirac equation where the velocity of light is replaced by the Fermi velocity~$v_F$.
Dirac cones are typically protected by symmetries, e.g., time-reversal and spatial inversion symmetry in the case of TBG~\cite{Manes2007,PhysRevX.8.031087,PhysRevX.8.031089}. 
Hence, upon tuning the bandwidth by a symmetry-compatible moir\'e potential, the Dirac cones continue to be present, however, with a modified Fermi velocity, which can become very small, e.g., close to the magic angle in TBG~\cite{Bistritzer2011,PhysRevX.8.031087}.

Here, we take these developments as a motivation to theoretically explore an analog gravity scenario for two-dimensional moir\'e Dirac materials in which the Fermi velocity can be tuned dynamically over several orders of magnitude.
In a geometric formulation, we show how to obtain an effectively time-dependent metric for the Dirac fermions.
In addition, we consider the presence of time-dependent Dirac masses that may originate from symmetry breaking in Dirac materials and lead to a finite band gap in the energy dispersion~\cite{PhysRevX.8.031089,PhysRevX.12.011061,lu2019superconductors,stepanov2020untying,PhysRevLett.123.157601,PhysRevX.8.031089,parthenios2023twisted}. 
These ingredients are relevant traits of highly tunable moir\'e Dirac materials, which allow us to construct an analog model for the phenomenon of cosmological fermion production in expanding universes, arising due to a time-dependent metric and conformal symmetry breaking \cite{Parker1971,Dolgov1995,Lyth1998,Giudice1999,Chung2000,Peloso2000,Greene2000,Asaka2010,Adshead2015,Scardua2018,Ema2019}. 

According to the modern understanding of cosmology, fermionic matter was created mainly during an epoch called reheating, following the early fast expansion referred to as inflation \cite{Kolb1990,Weinberg2008}. All fermionic matter that might have existed before inflation would be diluted during this rapid expansion so much that it is believed that the matter existing today must have been created afterwards.
The precise mechanism for fermionic matter production is at present not known, and it contains a number of mysteries. In particular, one would like to understand why more matter than anti-matter exists today, how precisely leptons and baryons have been created, and whether any physics beyond the standard model of elementary particle physics played a role for this.

Unfortunately, it is difficult to clarify these questions from astrophysical observations alone. Most of the information about the reheating phase has been lost during the evolution of the universe and a direct probe of the relevant high energy physics in laboratories (e.g., with particle colliders) also has limitations. This motivates us to explore here the possibilities to do quantum simulations, or analog experiments, with accessible condensed matter systems.

Producing relativistic fermions (usually as particle-anti-particle pairs) is possible by different mechanisms. One possibility would be inelastic collisions or reactions, as one could also imagine for classical particles in a kinetic description. But there are also other possibilities. For example, a coupling to an oscillating scalar field would make the effective Hamiltonian for the fermions time-dependent such that excitations get produced when the time dependence is strong enough to make the time evolution appreciably non-adiabatic (in the quantum-mechanical sense) \cite{Peloso2000,Chung2000,Asaka2010,Greene2000}. Time dependencies in the Hamiltonian can also appear for different reasons and another interesting possibility besides a time-dependent mass term is the time dependence of spacetime metric due to the cosmological expansion itself \cite{Kuzmin1999,Chung2012}.

On first sight one would think that particle production is most effective for massless relativistic fermions because there is then no energy gap to overcome. There is a subtlety here, however. Non-interacting, massless relativistic fermions are an example for a conformal quantum field theory, and such theories cannot be taken out-of-equilibrium by only an isotropic expansion of space as in an expanding universe. Only when conformal symmetry is broken, e.g., by a mass term (or energy gap in condensed matter notation) does the time-dependent cosmological metric by itself induce particle production \cite{Birrell1982,parker2009}.

In the condensed matter analogy, the effect of a time-dependent spacetime metric can be modeled by a time-dependent Fermi velocity $v_F$. However, by the above arguments, this only induces particle production together with a non-vanishing energy band gap $\Delta$. In fact, we find that it is the time dependence of the ratio $\Delta/v_F$ that matters. This implies that the effect we want to study could be induced either by a constant energy gap $\Delta$ with time-dependent Fermi velocity $v_F$, or a time-dependent gap $\Delta$ with constant Fermi velocity, or a combination thereof. We further exhibit that at non-zero temperature, when some modes have a thermal occupation, particle production is suppressed by Pauli blocking.
We discuss different options for realizing time-dependent $\Delta/v_F$ in our conclusion for several material platforms.

Interesting observables to confirm and further investigate fermionic quasiparticle production in Dirac materials are various two-point correlation functions. We therefore explore their time- and wavenumber-dependence in detail and, finally, we briefly discuss possibilities towards an experimental realization of the proposed quantum simulation.

The remainder of this work is organized as follows. In Sec.~\ref{sec:FermionicField}, we derive the spacetime metric corresponding to the low-energy fermionic excitations of a general Dirac material with dynamical Fermi velocity and we make a one-to-one mapping with an expanding Friedmann-Lema\^itre-Robertson-Walker (FLRW) metric. In Sec.~\ref{sec:ParticleProduction}, we discuss the phenomenon of quasiparticle production through the study of observables such as the occupation number or two-point correlation functions. In Sec.~\ref{sec:outlook}, we summarize and discuss further directions.

\paragraph*{Notation.} We work in SI units. For convenience, we drop the operator hats. The indices from the beginning of the Greek alphabet, $\alpha, \beta$, correspond to the constant spacetime running from $0$ to $2$, while the indices from the end, $\mu, \nu$, refer to the expanding spacetime coordinates $t, \vec{x}$. Also, vectors are denoted by bold symbols and Einstein summation convention is used.

\section{Fermionic field in a spatially flat Dirac material}\label{sec:FermionicField}

The aim of this chapter is to establish the spacetime metric formulation of a dynamically tunable, two-dimensional, but spatially flat Dirac material, e.g., twisted bilayer graphene, and to find a correspondence with the spatially flat Friedmann-Lema\^itre-Robertson-Walker (FLRW) metric \cite{Friedman1922,Friedman1924,Lemaitre1931,Robertson1935,Robertson1936a,Robertson1936b}. 
The FLRW metric provides a mathematical framework for modeling the geometry and dynamics of the universe on cosmological scales. 
It assumes an expanding (or contracting), homogeneous and isotropic universe, meaning that, on large scales, the universe appears the same in all directions and at all points. 
This is a fundamental assumption in cosmology, supported by observational evidence, such as the isotropic distribution of galaxies and the cosmic microwave background radiation.

\subsection{Dirac material metric and graphene representation}

Generally, the dynamics of Dirac fermions in a two-dimensional Dirac material can be described by the action $S[\Psi,\bar{\Psi}] =\int \text{d}t \, \text{d}^2 x \,  \mathcal{L}$ with Lagrangian density 
\begin{equation}
\begin{split}
     \mathcal{L}[\Psi,\bar{\Psi}]=-  \bar{\Psi}\left[\hbar \gamma^{0} \partial_{t} + \hbar v_F(t) \Vec{\gamma}\cdot\Vec{\partial} + \Delta_j(t) \Gamma^j \right]\Psi\,,
\label{eq:GrapheneLagrangian}
\end{split}
\end{equation}
where $\Psi$ is a four-component Dirac spinor, describing low-energy fermionic excitations near the Fermi surface, and $\bar{\Psi}=i\Psi^\dagger \gamma^0$ is the Dirac adjoint.
The Fermi velocity~$v_F$ will be considered as being time-dependent in the following and 
the spatial part of the Lagrangian contains the ${4\times 4}$~matrices $\Vec{\gamma}=(\gamma^1,\gamma^2)$, which are typically constructed from an underlying lattice-hopping model.
It is convenient to choose the basis of the Dirac field as 
\begin{equation}
    \Psi  = \begin{pmatrix}
    \psi^{+A} \\
    \psi^{+B} \\
    \psi^{-A} \\
    \psi^{- B}
    \end{pmatrix}\,,
\label{eq:ExpansionFieldDiracPointsSublattices}
\end{equation}
where the superscripts $A$ and $B$ may denote different sublattices of the considered Dirac material and~$+$ and~$-$ may label the two points $K$ and $K^\prime$ in momentum space where Dirac cones appear as, e.g., in the case of spinless fermions hopping on a honeycomb lattice~\cite{Katsnelson2012,Haldane1988,Herbut2006,Herbut2009,Boyack:2020xpe}.  We note that moir\'e Dirac materials can also have larger spinors or have components of a different origin, e.g., from layer, spin, or orbital degrees of freedom~\cite{Angeli_2021,PhysRevX.8.031089,Christos2020}. Here, we consider the four-component case as a minimal paradigmatic model and do not further restrict the microscopic origin of the indices $A,B$ and $\pm$.

The gamma matrices generate a  Clifford algebra in Minkowski spacetime, $\left\{\gamma^\alpha,\gamma^\beta\right\} = 2\eta^{\alpha\beta}$, where $\eta^{\alpha\beta}$ is the Minkowski metric with signature $(-,+,+)$. 
The time-like gamma matrix can be chosen as $\gamma^0 = i \, I_2 \otimes \sigma^3$ and the space-like gamma matrices as $\gamma^1 = \sigma^3 \otimes \sigma^2$ and ${\gamma^2 = - I_2 \otimes \sigma^1}$, with $I_2$ being the $2\times2$ identity matrix and $\vec{\sigma}$ the standard Pauli matrices \cite{Herbut2006,Herbut2009}.
The physical results do not depend on the explicit choice of the representation. 
The two remaining anticommuting gamma matrices can be taken as $\gamma^3 = \sigma^1\otimes \sigma^2  $ and $ \gamma^5 = i \gamma^0 \gamma^1 \gamma^2 \gamma^3 = \sigma^2\otimes \sigma^2$. 
Further, we also introduce $\gamma^{35}=\gamma^3\gamma^5=i\sigma^3\otimes I_2$. This specific choice of gamma matrices is often referred to as the graphene representation. 

In addition, we consider energy band gap openings at the Dirac points of the Dirac material. 
To this end, a linear superposition of mass terms in the Lagrangian density~\eqref{eq:GrapheneLagrangian} is included,
where $\Delta_j$ represents the amplitude of the mass gap and can be time-dependent, and the tensor $\Gamma^j$ depends on the shape of the gap. 
Such masses can be caused by symmetry breaking, e.g., when a sample of TBG is suspended on a substrate of hexagonal boron nitride (h-BN), a band gap is opened due to breaking of $C_2$ symmetry \cite{Cea2020,Long2022,Long2023}.
Another possible source can be the formation of strongly-correlated states and spontaneous symmetry breaking~\cite{PhysRevX.12.011061,lu2019superconductors,stepanov2020untying,PhysRevLett.123.157601,Herbut2006,Herbut2009,Boyack:2020xpe,PhysRevX.8.031089,parthenios2023twisted}.
Here, we exclusively examine masses that preserve Lorentz symmetry. This approach is motivated by our aim to establish a connection with cosmological scenarios, where Lorentz invariance is typically realized at the level of the Lagrangian. Therefore, in a (2+1)-dimensional spacetime the possible masses are proportional to ${\Gamma^j \in \{ I_4 , i \gamma^3 , i  \gamma^5 , i \gamma^{35} \} }$ and combinations of them.
However, Lorentz breaking gap terms can be interesting as well from the condensed matter perspective and could be studied further in the future.

For illustration, we briefly discuss the meaning of the different mass tensors for the paradigmatic case of Dirac excitations originating from spinless fermions hopping on a honeycomb lattice~\cite{Herbut2009}:
A mass term proportional to the identity matrix corresponds to the breaking of spatial inversion symmetry. 
Mass terms proportional to $\gamma^3$ and $\gamma^5$ also break spatial inversion and imply a linear mixing between the two Lorentz group irreducible representations, cf. Appendix~\ref{app:rep} for more details. 
Their combination $(\gamma^3\cos\alpha + \gamma^5\sin\alpha)$ can be understood as a Kekulé modulation of the nearest neighbor hopping. 
Finally, a mass term proportional to $\gamma^{35}$ corresponds to a Haldane mass, preserving handedness but breaking time-reversal symmetry \cite{Haldane1988,Herbut2009}. 
As a result of time-reversal symmetry breaking, a quantum Hall effect is generated without need of an external magnetic field when this kind of mass term is considered.
In the extended context of moir\'e Dirac materials, the meaning of the different mass gap contributions may be changed as compared to the case of spinless fermions on a honeycomb lattice, due to the different roles of the spinor components.

The general case allowed by Lorentz symmetry corresponds to a linear superposition of the different gap terms in Eq.~\eqref{eq:GrapheneLagrangian}. 
This band gap contribution in the Lagrangian for the Dirac material is related to a fermionic mass term in the cosmological Lagrangian density in an FLRW expanding universe. 
This contribution is essential for the phenomenon of particle production.

\subsection{Dirac fermions in curved spacetime}
For the following discussion it is helpful to recall briefly how Dirac fermion fields are described in general spacetimes, possibly curved. We concentrate on non-interacting fermions but allow for a general mass gap.

The Lagrangian density is of the form \cite{Parker1969,Birrell1982,Vozmediano2010,parker2009}
\begin{equation}
    \mathcal{L}_\text{D}[\Psi,\bar{\Psi}] = - \sqrt{g} \bar{\Psi} \left[\hbar\gamma^{\alpha} e_\alpha^{\phantom{\alpha}\mu} \left(\partial_{\mu} + \Omega_\mu\right) + m_j \Gamma^j \right]\Psi\,.
    \label{eq:LagrandianDirac}
\end{equation}
Different geometric fields enter here besides the actual Dirac spinor field $\Psi(x)$. One is $e_\alpha^{\phantom{\alpha}\mu}(x)$, the inverse of the vielbein $e^\alpha_{\phantom{\alpha}\mu}(x)$, which can be seen as a differential one-form, $e^\alpha_{\phantom{\alpha}\mu}(x) dx^\mu$, labeled by the index $\alpha$. The vielbein established a connection between a local orthonormal frame (with metric $\eta_{\alpha\beta}$) and the coordinate frame where the metric is 
\begin{equation}
    g_{\mu \nu} (x) =e^\alpha_{\phantom{\alpha}\mu}  (x) e^\beta_{\phantom{\beta}\nu} (x) \eta_{\alpha\beta}\,.
\end{equation}
The orthonormal frame is where the Clifford algebra with $\{\gamma^\alpha, \gamma^\beta\} = 2 \eta^{\alpha\beta}$ is situated. We also use the abbreviation for the determinant of the metric $g=-\det(g_{\mu\nu}(x))$.

Besides the vielbein field we are also using the spin connection $\Omega_{\mu}(x) = \omega_{\mu\alpha\beta}(x) \left[\gamma^\alpha,\gamma^\beta\right]/8$, in the appropriate representation for Dirac fermions. One can see the spin connection coefficient $\omega_{\mu\alpha\beta}(x)$ as a gauge field for local Lorentz transformations. It can be expressed through
\begin{equation}
    \omega_{\mu\alpha\beta}(x) = -\eta_{\alpha\gamma}\left[\partial_\mu e^\gamma_{\phantom{\gamma}\nu}(x) - \Gamma^\rho_{\phantom{\rho}\mu\nu}(x) e^\gamma_{\phantom{\gamma}\rho}(x) \right] e_\beta^{\phantom{\beta}\nu}(x)\,,
\end{equation}
in terms of the vielbein field and the connection in the metric frame $\Gamma^\rho_{\phantom{\rho}\mu\nu}(x)$. The latter is the Levi-Civita connection (see Appendix~\ref{app:General}) when torsion and non-metricity are absent.

The action in \eqref{eq:LagrandianDirac} is invariant under general coordinate changes (diffeomorphisms) and local Lorentz transformations. In addition to this, one can introduce another spacetime related transformation that will be very useful in the following: local Weyl scaling transformations. Here, the Dirac field transform with their scaling dimension $\Delta_\Psi=(d-1)/2=1$ in $d=1+2$ spacetime dimensions as
\begin{equation}
\begin{split}
   & \Psi(x) \to e^{-\Delta_\Psi \zeta(x)} \Psi(x), \\
    &    \bar\Psi(x) \to  e^{-\Delta_\Psi \zeta(x)} \bar\Psi(x)\,.
\label{eq:DefWeylScaling}
\end{split}
\end{equation}
At the same time the vielbein gets transformed like $e^\alpha_{\phantom{\alpha}\mu}(x) \to e^{\zeta(x)} e^\alpha_{\phantom{\alpha}\mu}(x)$ and the metric in the coordinate frame accordingly like $g_{\mu\nu}(x) \to \tilde{g}_{\mu\nu}(x) = e^{2\zeta(x)} g_{\mu\nu}(x)$. For the spin connection one has \cite{parker2009}
\begin{equation}
    \omega_{\mu\alpha\beta} \to \omega_{\mu\alpha\beta} + \left[e_{\alpha \mu} e_\beta^{\phantom{\beta}\nu} - e_{\beta\mu} e_\alpha^{\phantom{\alpha}\nu} \right]\partial_\nu \zeta\,.
\end{equation}
Using $\gamma_\alpha [\gamma^\alpha, \gamma^\beta] = 2(d-1)\gamma^\beta$ one can see that the Lagrangian in \eqref{eq:LagrandianDirac} is indeed invariant under local Weyl scaling transformations in the massless case, $m_j\Gamma^j=0$. If this gap term is non-vanishing, a Weyl transformation changes effectively
\begin{equation}
    m_j \Gamma^j \to e^{\zeta(x)} m_j\Gamma^j\,.
\end{equation}
In other words, a previously constant mass parameter becomes after the Weyl scaling in general space- and time-dependent. On the other side, one can use a Weyl scaling to go from a reference frame where the gap parameter depends in a factorizable way on time or space to another reference system where it is constant. Such a transformation changes the spacetime metric $g_{\mu\nu}(x)$ and the spin connection $\Omega_\mu(x)$. It is in this sense that a non-trivial spacetime metric and a time- or space-dependent gap parameter are two sides of the same coin.

\subsection{From Dirac material to an expanding universe}

Low-energy fermionic excitations in Dirac materials in the vicinity of Dirac cones have the characteristic linear dispersion relation of quasirelativistic fermions. 
These excitations propagate within the spacetime geometry set by a line element which is governed by a potentially time-dependent Fermi velocity.

We can now compare the Lagrangian density in \eqref{eq:GrapheneLagrangian} with time-dependent Fermi velocity $v_F(t)$ and gap parameter $\Delta_j(t)$  to the Lagrangian density for a Dirac field in curved spacetime in Eq.\ \eqref{eq:LagrandianDirac}. We find an agreement of the two theories for the choice 
\begin{equation}
    \left(e_\alpha^{\phantom{\alpha}\mu} \right)= \text{diag}\left( 1/v_F(t),1,1\right)\,.
    \label{eq:vielbeinFieldsCartesian}
\end{equation}
One finds therefore in this frame the effective metric of a Dirac material to be ${\left(g_{\mu \nu}\right)= \text{diag}\left( -v_F^2(t),1,1\right)}$, with determinant $\sqrt{g}  = v_F(t)$. A small calculation (see Appendix~\ref{app:WeylTransformation}) reveals that the spin connection $\Omega_\mu$ indeed vanishes in this case. 
One also has to identify
\begin{equation}
 m_j=\Delta_j(t)/v_F(t)\,.
 \label{eq:mjasRatio}
\end{equation}
Let us initially assume that $m_j$ is constant. Despite first appearance, this corresponds to a flat spacetime, as it becomes clear when a new time variable $\eta$ is chosen such that $\text{d}\eta = v_F(t) \text{d}t$, and the invariant spacetime element is simply $\text{d}s^2= - \text{d}\eta^2 + \text{d} \mathbf{x}^2$. 
Accordingly, in the case that the ratio on the right hand side of \eqref{eq:mjasRatio} is independent of time, we do not expect curved spacetime effects like particle production.

However, in general, the ratio on the right hand side of Eq.~\eqref{eq:mjasRatio} is \textit{not} independent of time. In this case, one can do a time-dependent Weyl scaling as specified in Eq.~\eqref{eq:DefWeylScaling}. With the choice
\begin{equation}
e^{\zeta(t)} = \frac{\Delta_j(0) v_F(t)}{\Delta_j(t) v_F(0)}\,,
\label{eq:ConformalFactor}
\end{equation}
after the transformation, we find then a constant mass term
\begin{equation}
    e^{\zeta(t)} \frac{\Delta_j(t) \Gamma^j}{v_F(t)} = m_j \Gamma^j\,.
\end{equation}
Note that this assumes that $\Delta_j(t)$ is non-vanishing only for one index $j$ or, more generally, that the time-dependence can be factored out. This Weyl scaling has changed  the spin connection, the vielbein fields and the metric,
\begin{equation}
    \begin{split}
        \text{d} s^2 
       &= \Tilde{g}_{\mu \nu} (x) \text{d} x^\mu \text{d} x^\nu\\  
&= e^{2\zeta(\eta)} \left[ - \text{d} \eta^2 + \text{d} \vec{x}^2 \right]\,,
    \end{split}
    \label{eq:GrapheneLineElement}
\end{equation}
and now they correspond to the case of curved spacetimes, see Appendix~\ref{app:WeylTransformation} for further details. 

Thus, considering a time-evolving Fermi velocity $v_F(\eta)$ and gap $\Delta(\eta)$, one can introduce the time-dependent scale factor $a(\eta)$, establishing a connection with cosmology,
\begin{equation}
    a(\eta) \equiv e^{\zeta(\eta)}\,,
\end{equation}
and reshape the former Dirac material line element into an FLRW line element in conformal time $\eta$
\begin{equation}
    \begin{split}
        \text{d} s^2 
        &=  a^2(\eta) \left[-\text{d}\eta^2 + \text{d}\Vec{x}^2\right]\,,
    \end{split}
    \label{eq:FLRWLineElement}
\end{equation}
with $a(\eta) \text{d}\eta=\text{d}t$.
In a cosmological context, the scale factor $a(\eta)$ governs the expansion of an FLRW universe and, hence, increasing (decreasing) the ratio between the Fermi velocity and the gap, $v_F(\eta)/\Delta(\eta)$, corresponds to a Dirac material analog of an expanding (contracting) universe.

\section{Particle production}
\label{sec:ParticleProduction}

We now address the phenomenon of fermionic pair production, which arises when a band gap in terms of a Dirac mass is considered and a time-dependent perturbation manifests through the Fermi velocity and/or the band gap.

\subsection{Dirac equation and mode functions}\label{sec:DiracEquationModeFunctions}

The generally covariant equation of motion for a Dirac field $\Psi$ in a spacetime background given by the time-dependent metric \eqref{eq:vielbeinFieldsCartesian} reads
\begin{equation}
\begin{split}
    0=\left[\hbar \gamma^0\partial_\eta +  \hbar  \vec{\gamma} \cdot \vec{\partial} + \frac{\Delta_j (\eta)}{v_F(\eta)} \Gamma^j \right]\Psi\,,
    \label{eq:DiracEquationMass}
\end{split}
\end{equation}
which is the well-known massive Dirac equation.

The Dirac field $\Psi$ obeys the equal-time canonical anticommutation relations. Within the graphene representation, the Dirac field satisfies 
\begin{equation}
     \left\{\Psi_{a}(t,\vec{x}),\Psi^{\dagger}_{b}(t,\vec{x'})\right\} =  \delta^{(2)}(\vec{x}-\vec{x'}) \delta_{ab}\,,
\label{eq:AnticommutationRelationFieldGraphene}
\end{equation}
where $a, b$ refer to the spinor indices~\cite{parker2009}, cf.~Eq.~\eqref{eq:AnticommutatorDiracField} in Appendix~\ref{app:General} for more details.

In flat space, it is convenient to expand the components of the Dirac field, cf. Eq.~\eqref{eq:ExpansionFieldDiracPointsSublattices}, in Fourier modes
\begin{equation}
     \psi^{\xi\lambda}(\eta,\vec{x}) 
    = \int^\Lambda \frac{\text{d}^2q}{(2\pi)^2}   \psi^{\xi\lambda}_{\vec{q}}(\eta) e^{i\vec{q}\cdot\vec{x}}\,,
    \label{eq:FourierExpansion}
\end{equation}
where the subscripts $\xi=\{+,-\}$ and $\lambda=\{A,B\}$ denote the respective spinor indices. 
 For our theory, we exclusively consider the quasirelativistic linear part of the energy bands, i.e. we do not take into account any bending of the bands that appears at some energy further away from the Fermi level.
Therefore, the UV cutoff $\Lambda$ should be adapted to the extent of the Dirac cone, e.g. for moir\'e Dirac materials $\Lambda\simeq1/a_M$  with $a_M$ being the moir\'e lattice constant.
For convenience, we suppress~$\Lambda$ in all expressions unless explicitly needed.

The Grassmann fields $\psi^{\xi\lambda}_{\vec{q}}$ may be expanded as~\cite{Birrell1982}
\begin{equation}
\psi^{\xi\lambda}_{\vec{q}}(\eta) 
= 
    u^{\xi\lambda}_{\Vec{q}}(\eta)
    c^{\xi}_{\vec{q}} +  
     v^{\xi\lambda}_{\Vec{-q}}(\eta)  
    d^{\xi\dagger}_{\vec{-q}}\, ,
    \label{eq:ExpansionDiracField}
\end{equation}
where we introduce annihilation (creation) operators for fermionic $c_{\vec{q}}^{\xi(\dagger)}$ and antifermionic $d_{\vec{q}}^{\xi(\dagger)}$ Dirac quasiparticles. 
They satisfy the fermionic anticommutation relations
\begin{equation}
\begin{split}
    \left\{c^{\xi }_{\vec{q}},c_{\vec{q'}}^{\xi^\prime \dagger}\right\} =\left\{d^\xi_{\vec{q}},d_{\vec{q'}}^{\xi^\prime\dagger}\right\}  = (2\pi)^2\, \delta^{(2)}(\vec{q}-\vec{q'})\,\delta^{\xi\xi^\prime}\,,
    \label{eq:AnticommutationsFermionAntifermionOperators}
\end{split}
\end{equation}
with all the other anticommutators equal to zero so that Eq.~\eqref{eq:AnticommutationRelationFieldGraphene} is satisfied.

In addition, we introduced the mode functions for fermions, $u^{\xi\lambda }_{\Vec{q}}(t)$, and antifermions, $v^{\xi\lambda}_{\Vec{q}}(t)$, which are solutions of the Dirac equation~\eqref{eq:DiracEquationMass} and fully contain the time dependence of the Dirac field. 
They are constrained by the canonical anticommutation relations of the Dirac field and by charge conjugation symmetry, as we will discuss in the next section.

\subsection{Symmetry transformations and mode functions}\label{sec:SymmetriesModeFunctions}
 
In Dirac materials, including the case of spinless fermions on a honeycomb lattice and TBG, Dirac cones are typically protected by time-reversal and spatial inversion symmetry and they are robust as long as these two fundamental discrete symmetries are obeyed.
For example, for the case of spinless fermions on a honeycomb lattice, the time-reversal operator interchanges the Dirac points leaving the sublattices~$A$ and~$B$ invariant~\cite{Manes2007,Katsnelson2012}
\begin{equation}
    \mathcal{T} \psi^{\xi\lambda} =  \psi^{\xi^\prime \lambda}\,,
\end{equation}
with $\xi\neq\xi^\prime$ and $\mathcal{T} = \sigma^1\otimes I_2 = i\gamma^1\gamma^5$; while the spatial inversion goes one step further and, apart from reversing the Dirac points, it also switches the sublattices
\begin{equation}
    \mathcal{P} \psi^{\xi\lambda} = \psi^{\xi^\prime \lambda^\prime},
\end{equation}
where $\xi\neq\xi^\prime$ and $\lambda\neq\lambda^\prime$, and the spatial inversion matrix is written as $\mathcal{P} = \sigma^1\otimes \sigma^1 = \gamma^0\gamma^3$. For the case of other Dirac materials, these operators have to be adjusted, accordingly.
Both symmetries can be broken by certain mass terms leading to the opening of a band gap. 

A charge conjugation operator $\mathcal{C}$ is a complex linear unitary operator exchanging fermions with antifermions, satisfying $\mathcal{C}^{-1}\gamma^\mu \mathcal{C} = -\left(\gamma^{\mu }\right)^T$ and $\mathcal{C}^{-1}=\mathcal{C}^T=\mathcal{C}^\dagger=-\mathcal{C}$~\cite{Bjorken1964}. Consequently, it depends on the choice of the representation for the gamma matrices. Within the graphene representation, the charge conjugation operator can be chosen as $\mathcal{C} = i\gamma^0\gamma^2 C= i I_2 \otimes\sigma^2 C$. Here, $C$ transforms annihilation (creation) operators as
\begin{equation}
    Cc_{\Vec{q}}^{\xi(\dagger)} = d_{-\Vec{q}}^{\xi(\dagger)} , \quad Cd_{\Vec{q}}^{\xi(\dagger)} = c_{-\Vec{q}}^{\xi(\dagger)}\,.
\end{equation}
The charge conjugation operator acts on the field as
\begin{equation}
    \mathcal{C} \Psi (\eta,\vec{x}) = \Psi^*(\eta,\vec{x})\,,
\end{equation}
where the $*$ superscript on a multi-component quantum field denotes hermitian conjugation of each component of the field, but without transposing the components.

Charge conjugation invariance implies the following relations between the fermionic and antifermionic mode functions
\begin{equation}
     v^{\xi A}_{\Vec{q}} = u^{\xi B *}_{-\Vec{q}} \quad \text{and} \quad  v^{\xi B}_{\Vec{q}} = -u^{\xi A*}_{-\Vec{q}}\,,
\label{eq:ChargeConjugationRelation}
\end{equation}
reducing the number of independent mode functions per field by half.
Additionally, the mode functions are restricted by the equal-time canonical anticommutation relations of the Dirac field, Eq.~\eqref{eq:AnticommutationRelationFieldGraphene}, and by the fermionic anticommutation relations satisfied by the annihilation and creation operators, Eq.~\eqref{eq:AnticommutationsFermionAntifermionOperators}. This yields
\begin{equation}
   u^{\xi\lambda}_{\Vec{q}} u^{\xi\lambda^\prime *}_{\Vec{q}}  + v^{\xi\lambda}_{\Vec{q}} v^{\xi\lambda^\prime *}_{\Vec{q}}  =  \delta^{\lambda\lambda^\prime}\,.
   \label{eq:NormalizationAnticommutation}
\end{equation}
By using Eq.~\eqref{eq:ChargeConjugationRelation}, Eq.~\eqref{eq:NormalizationAnticommutation} becomes
\begin{equation}
\begin{split}
\sum_{\lambda} \left\lvert u^{\xi \lambda}_{\Vec{q}} \right\rvert^2 = \left\lvert u^{\xi A}_{\Vec{q}} \right\rvert^2 + \left\lvert u^{\xi B}_{\Vec{q}} \right\rvert^2  = 1\,,
   \label{eq:NormalizationAnticommutationWithChargeConjugation}
\end{split}
\end{equation}
reducing the number of independent mode functions per Dirac field to two.

\subsection{Mode equations}\label{sec:ModeEquations}

In a non-static situation, where the Fermi velocity and/or the gap are not constant, the Dirac field cannot be expanded in plane waves with respect to time.
Instead, we have to solve the corresponding Dirac equation~\eqref{eq:DiracEquationMass}
to find the time dependence of the mode functions.
Importantly, we observe that Eq.~\eqref{eq:DiracEquationMass}  depends only on the ratio $\Delta_j(\eta)/v_F(\eta)$ 
and only if this ratio exhibits a time dependence, non-trivial effects like pair production can appear.
In particular,  if the gap vanishes, i.e. $\Delta_j(\eta) = 0$, a time-dependent Fermi velocity $v_F$ would not influence Eq.~\eqref{eq:DiracEquationMass}. 
This can be interpreted as a consequence of conformal symmetry of massless Dirac fermions.

For mass terms anticommuting with the massless differential Dirac operator, which includes $\gamma^\alpha$ with $\alpha=0,1,2$, we require $(\Gamma^j)^2=-I_4$.
Correspondingly, for commuting mass terms, we need $(\Gamma^j)^2=I_4$.
Then, applying another Dirac differential operator to the Dirac equation~\eqref{eq:DiracEquationMass}, with the sign of the mass term changed only for commuting mass terms, leads to the second order differential equation
\begin{equation}
    0=\!\left[\partial_\eta^2\! -\!  \vec{\partial}^2 + \frac{ \Delta_j^2(\eta)}{\hbar^2 v^2_F(\eta)} -
 \frac{1}{\hbar} \partial_\eta \left(  \frac{  \Delta_j(\eta)}{v_F(\eta)}\right) \gamma^0 \Gamma^j \right]\!\Psi.
    \label{eq:SecondOrderRedefinedMassiveNonStaticDiracConformalGeneralGap}
\end{equation}
Gap terms proportional to block diagonal matrices, e.g., the identity or $\gamma^{35}$, do not mix the Dirac points. Consequently, the mode functions of the different valleys can be analyzed separately. 
Otherwise, the four components of the field have to be considered simultaneously.

To solve the Dirac equation~\eqref{eq:DiracEquationMass} we take into account the mode expansion of the Dirac field, cf. Eq.~\eqref{eq:ExpansionDiracField}, which leads to a coupled system of differential equations for the mode functions. 
For example, for the band gap proportional to the identity matrix, we obtain for a specific choice of valley $\xi=\pm1$,
\begin{equation}
\begin{split}
0=\begin{pmatrix}
 i \partial_\eta + \Delta/(\hbar v_F) & \xi q e^{- \xi i\varphi}\\
  -\xi q e^{ \xi i\varphi}&  -i \partial_\eta + \Delta/(\hbar v_F)
\end{pmatrix}
\begin{pmatrix}
 u^{\xi A}_{\Vec{q}}\\
 u^{\xi B}_{\Vec{q}}
\end{pmatrix}\,.
\label{eq:DiracEquationModeFunctions}
\end{split}
\end{equation}
Here, the momentum vector has been rewritten in polar coordinates $q_x + \xi i q_y = q e^{\xi i\varphi}$.
Equivalently, Eq.~\eqref{eq:SecondOrderRedefinedMassiveNonStaticDiracConformalGeneralGap} with a gap proportional to the identity matrix yields the uncoupled second-order differential equations
\begin{equation}
\begin{split}
    &0= \left[\partial_\eta^2 + \omega^2_q(\eta) \varmp
   \frac{i}{\hbar} \partial_\eta \left( \frac{ \Delta(\eta)}{ v_F(\eta)}\right) \right] u^{\xi A (B)}_{\Vec{q}}(\eta)\,,
     \label{eq:ConformalModeEquationsMI}
\end{split}
\end{equation}
where $\omega^2_q(\eta) = q^2 + \Delta^2(\eta) /(\hbar v_F(\eta))^2$ and the sign is fixed through the diagonal values of $\gamma^0$ entering the last term in Eq.~\eqref{eq:SecondOrderRedefinedMassiveNonStaticDiracConformalGeneralGap}.

We note that Eq.~\eqref{eq:ConformalModeEquationsMI} corresponds to a differential harmonic oscillator equation with complex, time-dependent frequency. The mode functions $u^{\xi \lambda}_{\Vec{q}}$ can be determined by solving Eq.~\eqref{eq:DiracEquationModeFunctions} for a given set of initial conditions.

\subsection{Hamiltonian and time segments}

The Hamiltonian of the system is given by
\begin{equation}
\begin{split}
    H &= H_0 + H_\Delta=  \int \text{d}^2x \Bar{\Psi} \left(\hbar v_F \vec{\gamma}\cdot\vec{\partial} + \Delta_j \Gamma^j \right)\Psi,
  \label{eq:DiracHamiltonian}
\end{split}
\end{equation}
and, for a time-dependent $v_F$ and/or $\Delta$, it depends explicitly on time. 
Accordingly, the energy in excitations need not to be conserved.
The kinetic part of the Dirac Hamiltonian in Eq.~\eqref{eq:DiracHamiltonian} in the basis of Eq.~\eqref{eq:ExpansionDiracField} reads
\begin{equation}
\begin{split}
    H_0 = &-\hbar  v_F \int \frac{\text{d}^2 q}{(2\pi)^2} \sum_\xi \xi q \\
 &\times\Big\{2\text{Re}\left[e^{-\xi i\varphi} u^{\xi A*}_{\vec{q}} u^{\xi B}_{\vec{q}} \right]\left( c^{\xi \dagger}_{\vec{q}} c^{\xi }_{\vec{q}} - d^{\xi }_{\vec{q}} d^{\xi \dagger }_{\vec{q}}\right)\\
 &+\left[ e^{-\xi i\varphi} \left(u^{\xi B}_{\vec{q}}\right)^2 -e^{\xi i\varphi} \left( u^{\xi A}_{\vec{q}}\right)^2   \right] d^{\xi }_{-\vec{q}} c^{\xi }_{\vec{q}}\\
 &+\left[ e^{\xi i\varphi} \left(u^{\xi B*}_{\vec{q}}\right)^2  - e^{-\xi i\varphi} \left( u^{\xi A*}_{\vec{q}}\right)^2  \right] c^{\xi \dagger}_{\vec{q}} d^{\xi \dagger}_{-\vec{q}}  \Big\}.
\end{split}
\end{equation}
In the following, we exclusively study a specific mass term for Eq.~\eqref{eq:DiracHamiltonian}, i.e. the one proportional to the identity, $\Gamma^j =I_4$,
\begin{equation}
\begin{split}
    H_{\Delta} &= \Delta \int \text{d}^2x \bar{\Psi}\,  \Psi.
\end{split}
\end{equation}
Other forms of the mass gap are investigated in Appendix~\ref{app:GeneralBandGap}.
In the basis of Eq.~\eqref{eq:ExpansionDiracField} the massive part of Eq.~\eqref{eq:DiracHamiltonian} reads
\begin{equation}
\begin{split}
    H_{\Delta}
    =\,&  \Delta \! \int \frac{\text{d}^2 q}{(2\pi)^2} \sum_\xi\!\big[ \left(\abs{u^{\xi B}_{\vec{q}}}^2\! - 
    \!\abs{u^{\xi A}_{\vec{q}}}^2 \right) \left( c^{\xi\dagger}_{\vec{q}} c^\xi_{\vec{q}}\!-\!  d^{\xi}_{\vec{q}} d^{\xi\dagger}_{\vec{q}} \right)\\ 
    & - 2 u^{\xi A }_{\vec{q}} u^{\xi B}_{\vec{q}} d_{\vec{-q}}^{\xi} c_{\vec{q}}^{\xi} - 2 u^{\xi A *}_{\vec{q}} u^{\xi B*}_{\vec{q}} c_{\vec{q}}^{\xi\dagger} d_{\vec{-q}}^{\xi\dagger} \big].\notag
\end{split}
\end{equation}
The full Dirac Hamiltonian then reads~\cite{Giudice1999,Chung2000,Peloso2000}
\begin{equation}
\begin{split}
   H=&  \int \frac{\text{d}^2 q}{2\pi}  \sum_\xi  \big[ E_{\vec{q}}^\xi(\eta) \left( c^{\xi \dagger}_{\vec{q}} c^{\xi }_{\vec{q}} - d^{\xi }_{\vec{q}} d^{\xi \dagger }_{\vec{q}}\right)\\
   &+ F^\xi_{\vec{q}}(\eta) d^{\xi }_{-\vec{q}} c^{\xi }_{\vec{q}} + F^{\xi*}_{\vec{q}}(\eta) c^{\xi \dagger}_{\vec{q}} d^{\xi \dagger}_{-\vec{q}}  \big],
   \label{eq:FullHamiltonian}
\end{split}
\end{equation}
where
\begin{equation}
\begin{split}
    &E_{\vec{q}}^\xi = -2\xi \hbar v_F q \text{Re}\left[e^{-\xi i \varphi} u^{\xi A *}_{\vec{q}} u^{\xi B}_{\vec{q}}\right]\\
    &\phantom{E_{\vec{q}}^\xi =} +\Delta \left(1-2\left\lvert u^{\xi A}_{\vec{q}} \right\rvert^2\right),\\
    &F_{\vec{q}}^\xi = \xi\hbar v_F q \left[ e^{\xi i \varphi} \left(u^{\xi A}_{\vec{q}}\right)^2 - e^{-\xi i \varphi} \left(u^{\xi B}_{\vec{q}}\right)^2 \right] \\
    &\phantom{F_{\vec{q}}^\xi =}- 2 \Delta u^{\xi A}_{\vec{q}} u^{\xi B}_{\vec{q}},
    \label{eq:FunctionsHamiltonian}
\end{split}
\end{equation}
with $\left(E_{\vec{q}}^{\xi}\right)^2 + \abs{F_{\vec{q}}^\xi}^2 = \hbar^2 v_F^2 \omega_q^2$. 

\begin{figure}[t]
\centering
\includegraphics[width=0.9\linewidth]{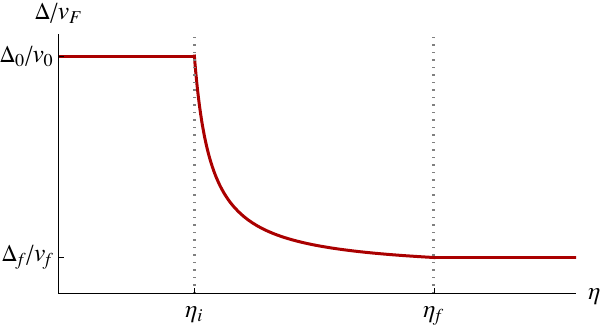}
\caption{\textbf{Time dependence of the ratio $\Delta(\eta)/v_F(\eta)$} for the three different temporal regions. In regions I $(\eta\leq\eta_{\text{i}})$ and III $(\eta\geq\eta_{\text{f}})$ the mass gap and the Fermi velocity are held constant, while in region II they become time-dependent. If their time dependence in region II is cancelled out, then no particles will be produced. The time dependencies shown in the plot are given by Eqs.~\eqref{eq:FermiVelocity} and \eqref{eq:Gap}.}
\label{fig:MassVelocity}
\end{figure}

We choose the Hamiltonian to be initially diagonal in terms of the set of operators $c^{\xi(\dagger)}_{\vec{q}}$ and $d^{\xi(\dagger)}_{\vec{q}}$ in a time segment where the Fermi velocity, $v_F(\eta\leq\eta_{\text{i}})\equiv v_0$, and the band gap, $\Delta(\eta\leq\eta_{\text{i}})\equiv\Delta_0$, are constant. 
We refer to this stationary time segment as region~I, cf. Fig.~\ref{fig:MassVelocity}.
The initial Hamiltonian~\eqref{eq:FullHamiltonian} becomes diagonal by taking a convenient initial configuration of mode functions corresponding to a no-particle state, leading to ${E_{\vec{q}}^\xi(\eta\leq\eta_\text{i}) = \hbar \omega_q^{\text{I}}}$ and $F_{\vec{q}}^\xi(\eta\leq\eta_\text{i}) = 0$.
Note that the last term of Eq.~\eqref{eq:SecondOrderRedefinedMassiveNonStaticDiracConformalGeneralGap} vanishes in stationary regions and it becomes a massive Klein-Gordon equation. 
Consequently, expanding the Dirac field in Fourier modes, the mode functions in this region are  solutions of a harmonic oscillator differential equation and can be taken as standard Minkowski modes.
Accordingly, a possible initial configuration of the mode functions, diagonalizing the initial Hamiltonian for a band gap proportional to the identity matrix, is given by
\begin{equation}
\begin{split}
   u^{\xi A, \text{I}}_{\vec{q}} (\eta\leq\eta_\text{i}) &= \frac{e^{-\xi i\varphi/2}}{\sqrt{2}}\sqrt{1 - \frac{ \Delta_0}{\hbar v_0 \omega^{\text{I}}_q}} e^{-i\omega^{\text{I}}_q \eta},\\
    u^{\xi B, \text{I}}_{\vec{q}} (\eta\leq\eta_\text{i}) &= -\xi \frac{e^{\xi i\varphi/2}}{\sqrt{2}}\sqrt{1 + \frac{ \Delta_0}{\hbar v_0 \omega^{\text{I}}_q}} e^{-i\omega^{\text{I}}_q \eta},
    \label{eq:InitialConditions}
\end{split}
\end{equation}
with positive frequency $\omega^{\text{I}}_q \equiv \omega_q (\eta_\text{i}) = \sqrt{q^2 + \Delta_0^2/(\hbar v_0)^2}$. 
With this, the behavior is indeed compatible with the one for Lorentz transformations, see Appendix~\ref{app:lorentz}.
In a similar way one can find for each band gap a set of initial mode functions corresponding to the no-particle state and with an initial Hamiltonian in the standard diagonal form. 
The (creation) annihilation fermionic $c^{\xi(\dagger)}_{\vec{q}}$ and antifermionic $d^{\xi(\dagger)}_{\vec{q}}$ operators associated to the mode functions $u^{\xi \lambda}_{\vec{q}}$ define an initial "$c$-" and "$d$-vacuum" state $\ket{\Omega}$  for such excitations,
\begin{equation}
    c^{\xi}_{\vec{q}}\ket{\Omega} = d^{\xi}_{\vec{q}}\ket{\Omega} = 0.
\end{equation}
Here, the vacuum state $\ket{\Omega}$ is a free state of excitations, which describes the ground state for a Fermi system with a filling up to a certain level. More generally, one can consider a different initial state with excitations, such as one characterized by a fixed temperature $T$.

Now, we assume that at time $\eta_\text{i}$ a dynamical time segment begins. There, the Fermi velocity $v_F(\eta)$ and/or the band gap $\Delta(\eta)$, but in particular their ratio $\Delta(\eta)/v_F(\eta)$, become time-dependent up to a time $\eta_\text{f}$. We refer to this time segment as region II in the following and the static regime following region II is region III, cf. Fig.~\ref{fig:MassVelocity}.
In this region, the mode functions are obtained by solving Eq.~\eqref{eq:DiracEquationMass} taking into consideration their initial conditions~\eqref{eq:InitialConditions}. 
The solution will not be the simple plane waves of Minkowski space, but a more involved function of time.

The functional forms of the mass gap $\Delta(\eta)$ and the Fermi velocity $v_F(\eta)$ for region II are chosen for simplicity as linear in the conformal time $\eta$,
\begin{equation}
    \Delta(\eta) = \Delta_0 -  D\left(\eta-\eta_\text{i}\right)\,,
    \label{eq:Gap}
\end{equation}
and
\begin{equation}
    v_F(\eta) = v_0 + V\left(\eta-\eta_\text{i}\right)\,,
    \label{eq:FermiVelocity}
\end{equation}
where $V$ and $D$ are two real, positive, constant factors.
Their ratio is shown in Fig.~\ref{fig:MassVelocity}. 
From the condensed matter point of view, tuning the ratio of the band gap and Fermi velocity in a real Dirac material may be achieved by different strategies, depending on the actual material system. For example, in a correlated moir\'e Dirac material, increasing the Fermi velocity, tentatively suppresses the effect of the interactions. 
Consequently, a decrease of the band gap can be expected. 
In other Dirac materials, it may even be possible to use an external field to modify the band gap.
However, note that the choice of time dependence is flexible and can be tailored to suit both experimental feasibility and scientific interest. 
While our work presents a specific choice of such time dependencies, it can be adjusted to align with the constraints of a real experimental setup.
Here, we choose their time-dependencies to be linear in conformal time for illustrative purposes, allowing us to present results on the characteristic signatures of fermion production in the Figs.~\ref{fig:BetaBogoliubov},\ref{fig:StatisticalTwoPointCorrelationFunction} and \ref{fig:TwoPointCorrelationFunctionsForG35}.
From a cosmological point of view, different time dependencies of the ratio can be chosen to study different cosmological scenarios, as $a(t)\propto v_F/\Delta$. 
For instance, $v_F/\Delta \propto e^{H t} $ corresponds to a de Sitter universe, $v_F/\Delta  \propto \abs{t-t_0}^{\gamma}$ with $\gamma=2/3$ to a radiation dominated universe and with $\gamma=1$ to a matter dominated universe.
Other interesting cosmologies can be studied by choosing a proper time dependence of this ratio.

After the time-dependent perturbation has ceased, i.e. for times $\eta>\eta_\text{f}$, the Hamiltonian of the system is not diagonal in the basis $c^{\xi(\dagger)}_{\vec{q}}$ and $d^{\xi(\dagger)}_{\vec{q}}$. Instead, a new set of creation and annihilation operators $\Tilde{c}^{\xi(\dagger)}_{\vec{q}}$ and $\Tilde{d}^{\xi(\dagger)}_{\vec{q}}$ can be introduced such that
\begin{equation}
    H = \hbar v_F \int \frac{\text{d}^2 q}{(2\pi)^2}  \omega_q \sum_\xi  \left( \tilde{c}^{\xi \dagger}_{\vec{q}} \tilde{c}^{\xi }_{\vec{q}} - \tilde{d}^{\xi }_{\vec{q}} \tilde{d}^{\xi \dagger }_{\vec{q}}\right).
    \label{eq:HamiltonianDiagonal}
\end{equation}
In this stationary region, we assume the Fermi velocity, $v_F(\eta\geq \eta_\text{f}) \equiv v_{\text{f}}$, and the band gap, $\Delta(\eta\geq \eta_\text{f})\equiv \Delta_\text{f}$, to be constant, again. 
A set of solutions for the mode functions can be found in terms of Minkowski modes with positive frequencies as for region I by solving Eq.~\eqref{eq:SecondOrderRedefinedMassiveNonStaticDiracConformalGeneralGap}. 
For a gap term proportional to the identity matrix one can choose
\begin{equation}
\begin{split}
    \Tilde{u}^{\xi A, \text{III}}_{\vec{q}}(\eta\geq \eta_\text{f}) &= \frac{e^{-\xi i \varphi/2}}{\sqrt{2}} \sqrt{1-\frac{\Delta_\text{f}}{\hbar v_{\text{f}} \omega_q^{\text{III}} } } e^{-i \omega_q^{\text{III}}\eta},\\
    \Tilde{u}^{\xi B, \text{III}}_{\vec{q}}(\eta\geq \eta_\text{f}) &= -\xi\frac{e^{\xi i \varphi/2}}{\sqrt{2}} \sqrt{1+\frac{\Delta_\text{f}}{\hbar v_{\text{f}} \omega_{q}^{\text{III}} } } e^{-i \omega_q^{\text{III}}\eta},
\end{split}
\end{equation}
where now $\omega_q^{\text{III}} \equiv \omega_q (\eta_\text{f}) = \sqrt{q^2 + \Delta_\text{f}^2/(\hbar v_\text{f})^2}$. 

\subsection{Bogoliubov transformation}

The new set of annihilation (creation) operators $\Tilde{c}^{\xi(\dagger)}_{\vec{q}}$ and $\Tilde{d}^{\xi(\dagger)}_{\vec{q}}$ associated to the mode functions $\Tilde{u}^{\xi \lambda}_{\vec{q}}$ that are Minkowski modes in region III define a new vacuum state $\ket{\Tilde{\Omega}}$ for such excitations,
\begin{equation}
    \Tilde{c}^{\xi}_{\vec{q}}\ket{\Tilde{\Omega}} = \Tilde{d}^{\xi}_{\vec{q}}\ket{\Tilde{\Omega}} = 0.
\end{equation}
The old mode functions $u^{\xi \lambda}_{\vec{q}}$ can be expressed in terms of the new set $\tilde{u}^{\xi \lambda}_{\vec{q}}$ through the Bogoliubov transformation
\begin{equation}   
    \begin{pmatrix}
        u^{\xi A}_{\vec{q}}\\
        u^{\xi B *}_{\vec{q}} 
    \end{pmatrix} 
    = \sum_\zeta \begin{pmatrix}
        \alpha^{\xi\zeta}_{\vec{q}} &  -\left(\beta^{\xi\zeta}_{\vec{q}}\right)^* \\
       \beta^{\xi\zeta}_{\vec{q}} & \left(\alpha^{\xi\zeta}_{\vec{q}}\right)^*
    \end{pmatrix}
    \begin{pmatrix}
        \Tilde{u}^{\zeta A}_{\vec{q}}\\
        \tilde{u}^{\zeta B *}_{\vec{q}} 
    \end{pmatrix}
\end{equation}
which, in region III, corresponds to a linear superposition of positive and negative frequency-mode solutions, with $\alpha_{\vec{q}}^{\zeta\xi}$ and $\beta_{\vec{q}}^{\zeta\xi}$ being complex-valued and time-independent Bogoliubov coefficients. 
Here, we have introduced a new superscript $\zeta$ for the new set of operators, which is not equivalent to $\xi$ when the mass term mixes the Dirac points. Otherwise, when the mass term is block diagonal as the one studied here, there are only four non-vanishing Bogoliubov coefficients matrix elements for $\zeta=\xi$. 

Therefore, the Bogoliubov transformation reduces to 
\begin{equation}
\begin{split}
    u^{\xi A}_{\vec{q}} =\,& \alpha^{\xi}_{\vec{q}} \Tilde{u}^{\xi A}_{\vec{q}} - \beta^{\xi *}_{\vec{q}} \Tilde{u}^{\xi B *}_{\vec{q}}\,,\\
    u^{\xi B}_{\vec{q}} =\,& \alpha^{\xi}_{\vec{q}} \Tilde{u}^{\xi B}_{\vec{q}} + \beta^{\xi *}_{\vec{q}}\Tilde{u}^{\xi A *}_{\vec{q}}\,,
\end{split}
\end{equation}
where the Bogoliubov coefficients are simplified by dropping one of the superscripts $\alpha^{\xi\xi}_{\vec{q}} \equiv \alpha^\xi_{\vec{q}}$ and $\beta^{\xi\xi}_{\vec{q}} \equiv \beta^\xi_{\vec{q}}$.

Furthermore, one can introduce the general Bogoliubov transformation that connects the two sets of creation and annihilation operators
\begin{equation}
    \begin{pmatrix}
        \Tilde{c}^\zeta_{\vec{q}}\\
        \tilde{d}^{\zeta\dagger}_{\vec{-q}} 
    \end{pmatrix} = \sum_\xi
    \begin{pmatrix}
        \alpha^{\zeta\xi}_{\vec{q}} & \beta^{\zeta\xi}_{\vec{q}} \\
        -\left(\beta^{\zeta\xi}_{\vec{q}}\right)^* & \left(\alpha^{\zeta\xi}_{\vec{q}}\right)^*
    \end{pmatrix}
    \begin{pmatrix}
        c^\xi_{\vec{q}}\\
        d^{\xi\dagger}_{\vec{-q}} 
    \end{pmatrix} 
    \label{eq:BogoliubovTransformationOperators}
\end{equation}
 which for a non-mixing mass term reduces to
\begin{equation}
\begin{split}
   \Tilde{c}^\xi_{\vec{q}} &= \alpha^{\xi}_{\vec{q}}c^{\xi}_{\vec{q}} + \beta^{\xi}_{\vec{q}} d^{\xi\dagger}_{\vec{-q}}, \\
  \Tilde{d}^{\xi\dagger}_{\vec{-q}} &=  -\beta^{\xi *}_{\vec{q}} c^{\xi}_{\vec{q}} + \alpha^{\xi *}_{\vec{q}} d^{\xi\dagger}_{\vec{-q}} .
    \label{eq:BogoliuvobTransformationOperators}
\end{split}
\end{equation}
The new set of operators also satisfies the canonical anticommutation relations in Eq.~\eqref{eq:AnticommutationsFermionAntifermionOperators} from which the following identities for the Bogoliubov coefficients matrix elements  can be derived
\begin{align}
   \sum_{\xi} \left(\alpha^{\zeta\xi}_{\vec{q}} \alpha^{ \zeta^\prime  \xi*}_{\vec{q}} +  \beta^{ \zeta  \xi}_{\vec{q}} \beta^{ \zeta^\prime \xi *  }_{\vec{q}}  \right)\equiv \delta^{\zeta\zeta^\prime}\,
\label{eq:BogoliubovCoefficientConditionGeneralGap}
\end{align}
and
\begin{align}
   \sum_{\xi} \left(\alpha^{\zeta\xi}_{\vec{q}} \beta^{ \zeta^\prime  \xi}_{\vec{q}} -  \beta^{ \zeta  \xi}_{\vec{q}} \alpha^{ \zeta^\prime \xi   }_{\vec{q}}  \right)\equiv 0\,.
\label{eq:BogoliubovCoefficientConditionGeneralGap2}
\end{align}
Eq.~\eqref{eq:BogoliubovCoefficientConditionGeneralGap} guarantees that the space density cannot exceed unity, as expected from the Pauli exclusion principle~\cite{Ema2019}. For a mass that does not mix the Dirac points the former identity reduces to ${\abs{\alpha_{\vec{q}}^\xi}^2+\abs{\beta_{\vec{q}}^\xi}^2 = 1}$ \cite{Lyth1998,Peloso2000}.

The Bogoliubov coefficients are obtained from diagonalizing the Hamiltonian after the dynamical process by performing a Bogoliubov transformation~\eqref{eq:BogoliuvobTransformationOperators} such that
\begin{equation}
\begin{split}
    \frac{\beta^\xi_{\vec{q}} }{\alpha^\xi_{\vec{q}}} = \frac{F_{\vec{q}}^{\xi *}}{\hbar   \omega_q + E_{\vec{q}}^\xi} \quad \text{and} \quad \abs{\beta^\xi_{\vec{q}}}^2 = \frac{\hbar  \omega_q - E_{\vec{q}}^\xi}{2\hbar  \omega_q}.
    \label{eq:BogoliubovCoefficients}
\end{split}
\end{equation}

\subsection{Number of excitations and correlation functions}\label{sec:Results}

In this section, we study observables that give signatures of quasiparticle production. These observables can be investigated starting from an initial vacuum state or from a state in which there are already fermionic excitations, e.g., a thermal state.

An initial state at $t_\text{i}$ with non-vanishing occupation number leads to an initial fermionic and antifermionic quasiparticle distribution given by 
\begin{equation}
\begin{split} N_{\text{in},c}^\xi (\vec{q}) =\braket{c_{\vec{q}}^{\xi\dagger}c_{\vec{q}}^\xi} \quad \text{and} \quad N_{\text{in},d}^\xi (\vec{q})= \braket{d_{\vec{q}}^{\xi\dagger} d_{\vec{q}}^\xi},
\end{split}
\end{equation}
respectively, and consequently, to Pauli blocking. 
Therefore, quasiparticle creation will be suppressed, when starting from an initial state that is different to the vacuum of excitations, as fermions cannot be created in states that are already occupied.
For instance, one can consider an initial thermal state,  in which the amount of fermions and antifermions is balanced, i.e. their initial occupation number will be the same, $N_{\text{in}}^\xi \equiv N_{\text{in},c}^\xi = N_{\text{in},d}^\xi $, following the Fermi-Dirac statistics
\begin{equation}
    N^\xi_{\text{in}}(T,q) = \frac{1}{e^{\hbar \omega^{\text{I}}_q/(k_\text{B} T)}+1},
    \label{eq:Thermal}
\end{equation}
with temperature~$T$ and Boltzmann constant~$k_\text{B}$.

\subsubsection{Expected number of quasiparticles}

\begin{figure}[t]
\centering
\includegraphics[height=0.75\linewidth]{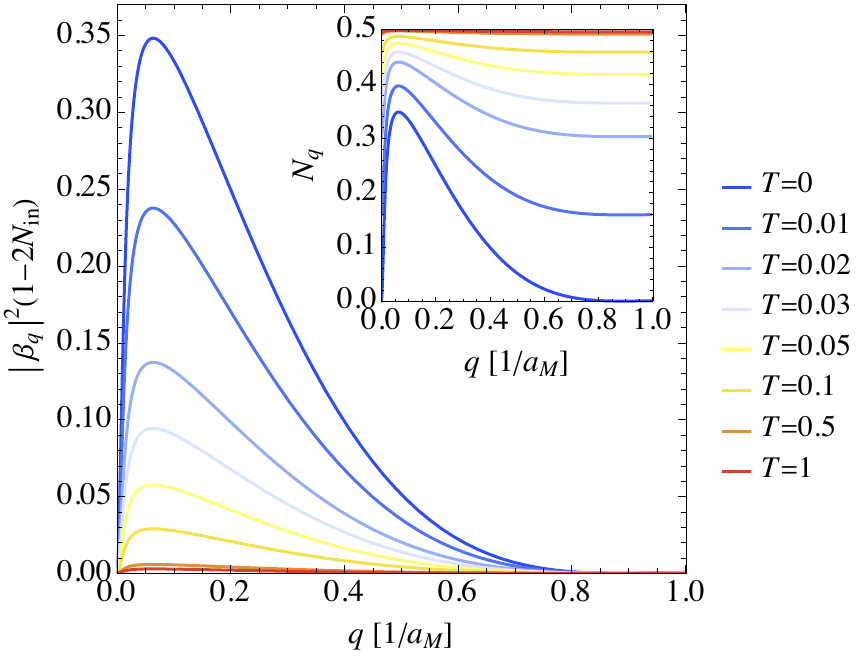}
\caption{\textbf{Spectrum of produced particles} including thermal Pauli blocking, cf. the last term of the expected number of quasiparticles in Eq.~\eqref{eq:ThermalOccupationNumber}. The inset shows the expected number of quasiparticles, $N_q$, of fermionic excitations in one Dirac point as a function of momentum $q$ in units of  $1/a_M$. This has been calculated for a band gap term proportional to the identity matrix starting with an initial no-particle state with $\beta_{\vec{q}}^{\text{in}}=0$.  
The same plot is obtained for the other non-symmetric Dirac point. The dynamical process (region~II) has a duration of $t_\text{f} - t_{\text{i}} = a_M/(5 v_0)$. The Fermi velocity $v_F$ is changed here from $v_0$ to $v_\text{f}=10^2 v_0$ and the band gap from $\Delta_0=50 \hbar v_0/a_M$ to $\Delta_\text{f}=\hbar v_0/a_M$ with the time dependence given in Eqs.~\eqref{eq:FermiVelocity} and~\eqref{eq:Gap}. The colors correspond to different thermal initial states with temperature $T$ in units of $\hbar v_0/(k_\text{B} a_M)$. The same plot is obtained for a gap proportional to $\gamma^{35}$ starting with its corresponding initial no-particle state. } 
\label{fig:BetaBogoliubov}
\end{figure}

To study the occupation number of Dirac fermion-antifermion pairs created per mode $q$, we evaluate the number operator of the new set of (creation) annihilation operators, i.e. $\tilde{c}^{\xi(\dagger)}_{\vec{q}}, \,\tilde{d}^{\xi(\dagger)}_{\vec{q}}$, at the initial vacuum state $\ket{\Omega}$
\begin{equation}
\begin{split}
   N^\xi_{\vec{q}} = \bra{\Omega} \tilde{c}^{\xi\dagger}_{\vec{q}} \tilde{c}^\xi_{\vec{q}} \ket{\Omega}
    = \abs{\beta^{\xi}_{\vec{q}}}^2 (2\pi)^2\delta^{(2)}(0),
    \label{eq:OccupationNumberVacuum}
\end{split}
\end{equation}
where the divergent factor $\delta^{(2)}(0)$ is a consequence of considering an infinite spatial volume, which arises from the anticommutation relation~\eqref{eq:AnticommutationsFermionAntifermionOperators} evaluated at equal momentum. 
The divergent factor can be substituted by considering a finite spatial volume $V$. 
From now on, we omit the factors of volume assuming that the number of particles are always referred to a unit of volume for simplicity of notation \cite{Calzetta2005,Mukhanov2007}.

As a consequence of charge conjugation symmetry, the expected occupation number of fermions and antifermions is the same.
Integrating over all momentum modes one gets the quasiparticle number density
\begin{equation}
\begin{split}
    n = \int  \frac{\text{d}^2 q}{(2\pi)^2} \sum_{\xi} N^\xi_{\vec{q}}.
\end{split}
\end{equation}
Taking an initial excited state instead of the vacuum or ground state, the expected number of quasiparticles is
\begin{equation}
\begin{split}
   N^\xi_{\vec{q}} = \braket{ \tilde{c}^{\xi\dagger}_{\vec{q}} \tilde{c}^\xi_{\vec{q}}}
    = N^\xi_{\text{in}} + \abs{\beta^{\xi}_{\vec{q}}}^2 \left(1- 2 N^\xi_{\text{in}} \right),
    \label{eq:ThermalOccupationNumber}
\end{split}
\end{equation}
where the last term corresponds to the Pauli blocking, shown in Fig.~\ref{fig:BetaBogoliubov} as well as the expected number of quasiparticles for one Dirac point in the inset. One can observe that the expected number of quasiparticles saturates at $1/2$ for all momentum modes for higher temperatures, as expected. The higher the temperature, the higher the initial number of excitations and, consequently, fewer quasiparticles are produced due to Pauli blocking and the main contribution to the expected number of quasiparticles comes from the initial number of excitations.

\subsubsection{Two-point correlation functions}\label{sec:TwoPointCorrelationFunctions}

\begin{figure*}
\centering
\begin{minipage}[t]{.48\textwidth}
\includegraphics[height=0.74\linewidth]{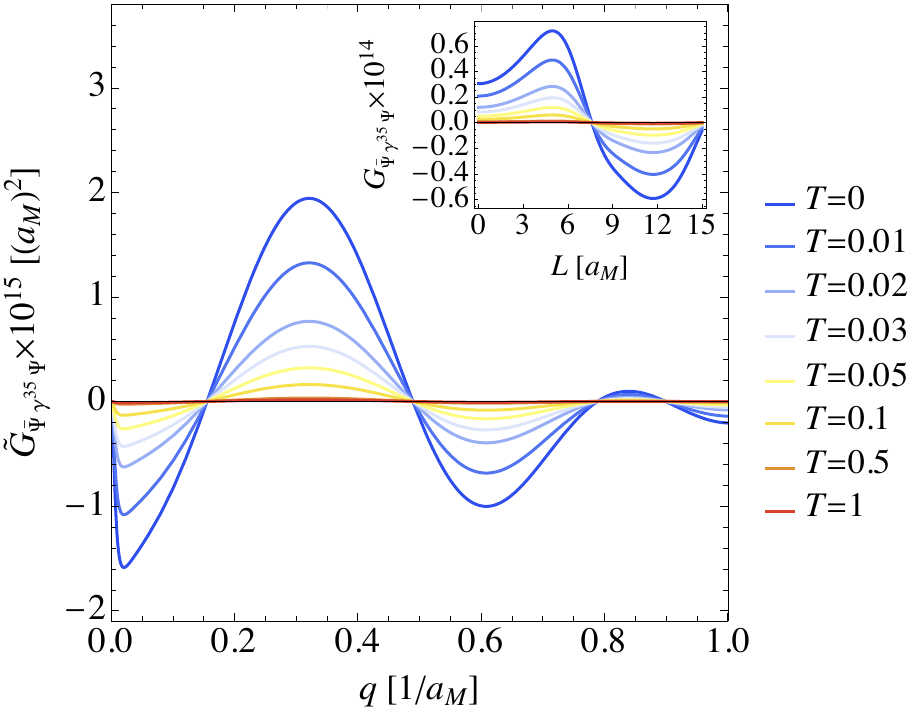}
\end{minipage}
\begin{minipage}[t]{.48\textwidth}
\includegraphics[height=0.74\linewidth]{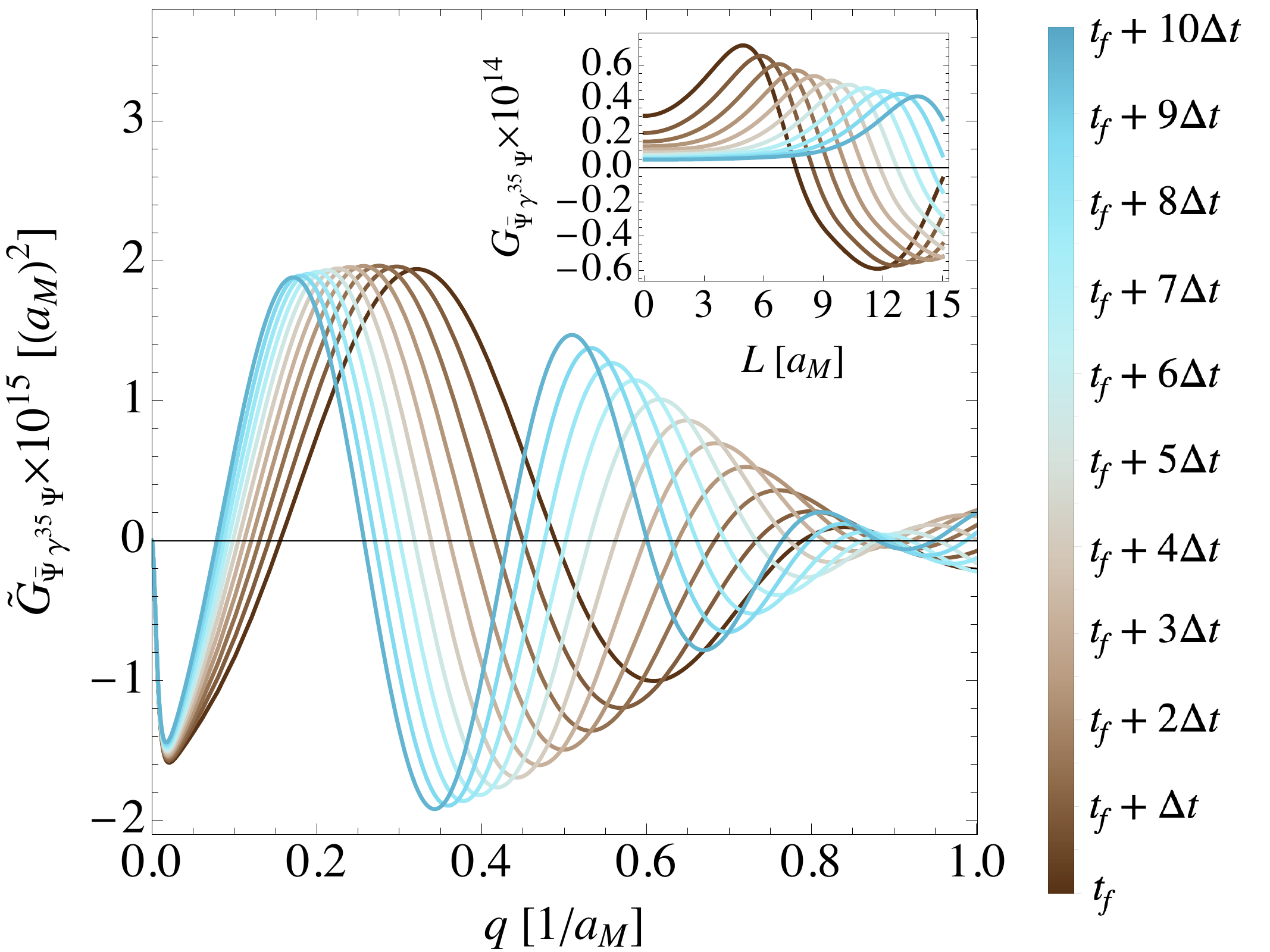}
\label{label-b}
\end{minipage}
\begin{minipage}[t]{.48\textwidth}
\includegraphics[height=0.74\linewidth]{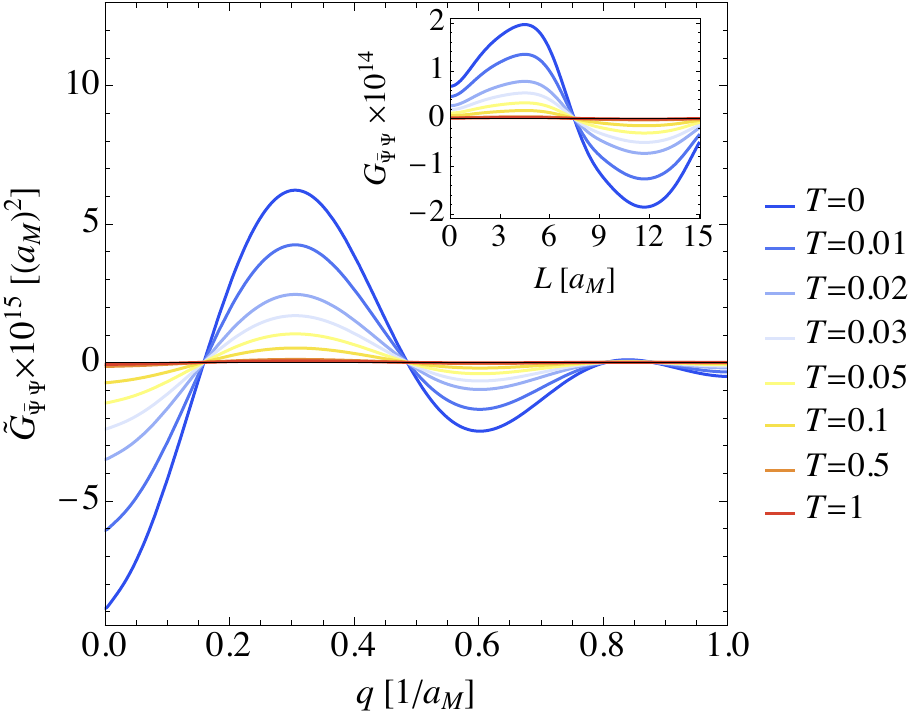}
\end{minipage}\quad
\begin{minipage}[t]{.48\textwidth}
\includegraphics[height=0.74\linewidth]{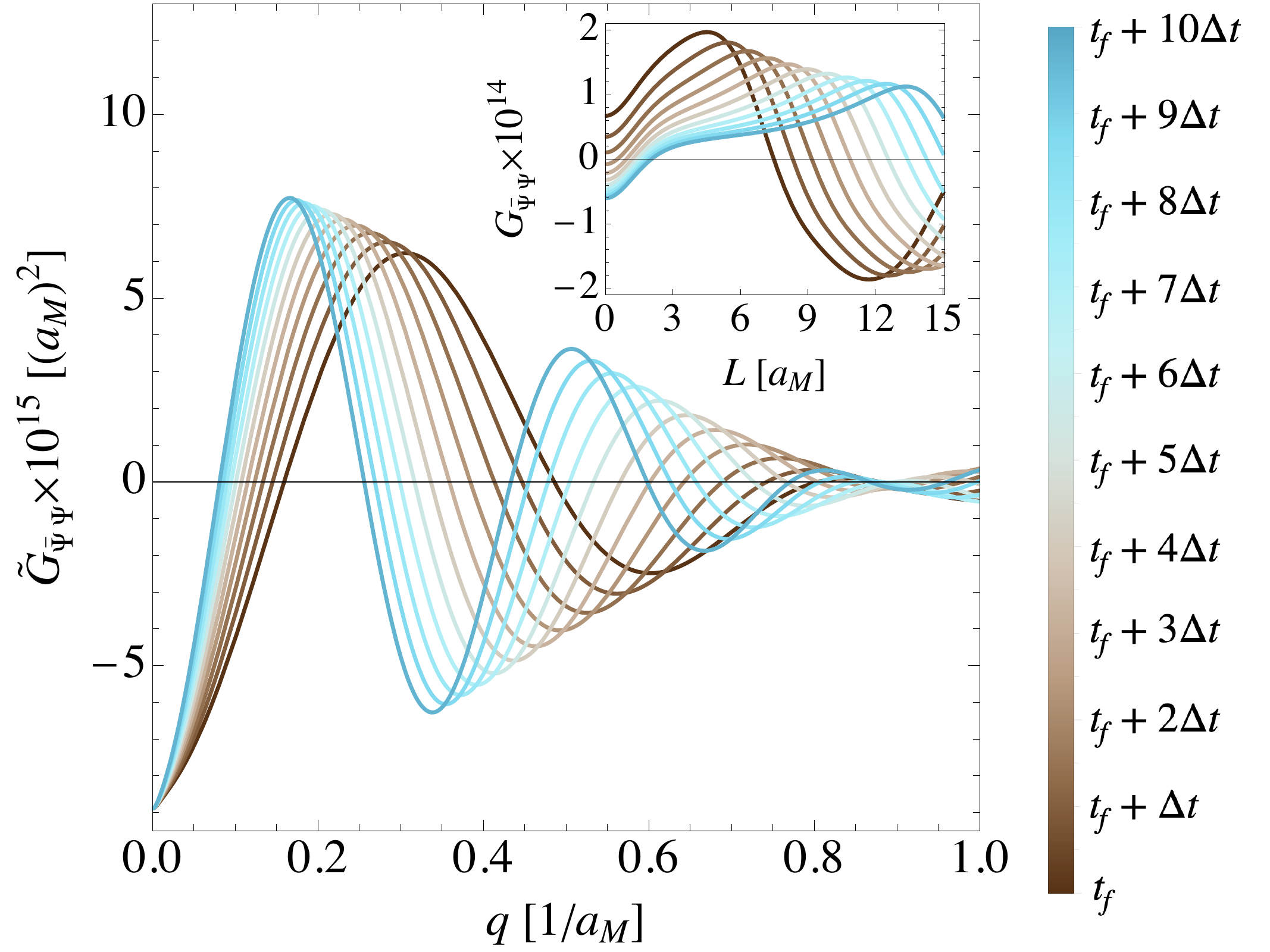}
\label{label-c}
\end{minipage}
\caption{\textbf{Equal-time two-point correlation functions of fermionic excitations.} From top to bottom: Equal-time two point correlation function of fermionic excitations in momentum space $\tilde{\mathcal{G}}^S_{\bar{\Psi}\gamma^{35}\Psi}$ and statistical equal-time two-point correlation function of fermionic excitations (also known as static structure factor) in momentum space $\tilde{\mathcal{G}}^S_{\bar{\Psi}\Psi}$ in units of~$(a_M)^{2}$ as a function of momentum~$q$ in units of~$1/a_M$. 
This has been calculated for a band gap term proportional to the identity matrix.  
The time dynamical process in region II has a duration of $t_\text{f} - t_{\text{i}} = a_M/(5 v_0)$. 
The Fermi velocity $v_F$ is changed here from $v_0$ to $v_\text{f}=10^2 v_0$ and the band gap from $\Delta_0=50 \hbar v_0/a_M$ to $\Delta_\text{f}=\hbar v_0/a_M$ with a time dependence given in Eqs.~\eqref{eq:FermiVelocity} and \eqref{eq:Gap}. 
In the inset the dimensionless equal-time two-point correlation functions in position space as a function of the distance $L$ are shown obtained after a regularization with a Gaussian window function of width $w=a_M$. 
 In the left panels, the colors correspond to different initial thermal states with temperature $T$ in units of $\hbar v_0/(k_\text{B} a_M)$. The two-point correlation functions are evaluated at final time $t_{\text{f}}$.  
In the right panels, the initial state is taken to be the vacuum at T=0K. The colors correspond to different holding times after the expansion has ceased with $\Delta t= a_M/(50 v_0)$.}
\label{fig:StatisticalTwoPointCorrelationFunction}
\end{figure*}

\begin{figure*}
\centering
\begin{minipage}[t]{.47\textwidth}
\includegraphics[height=0.77\linewidth]{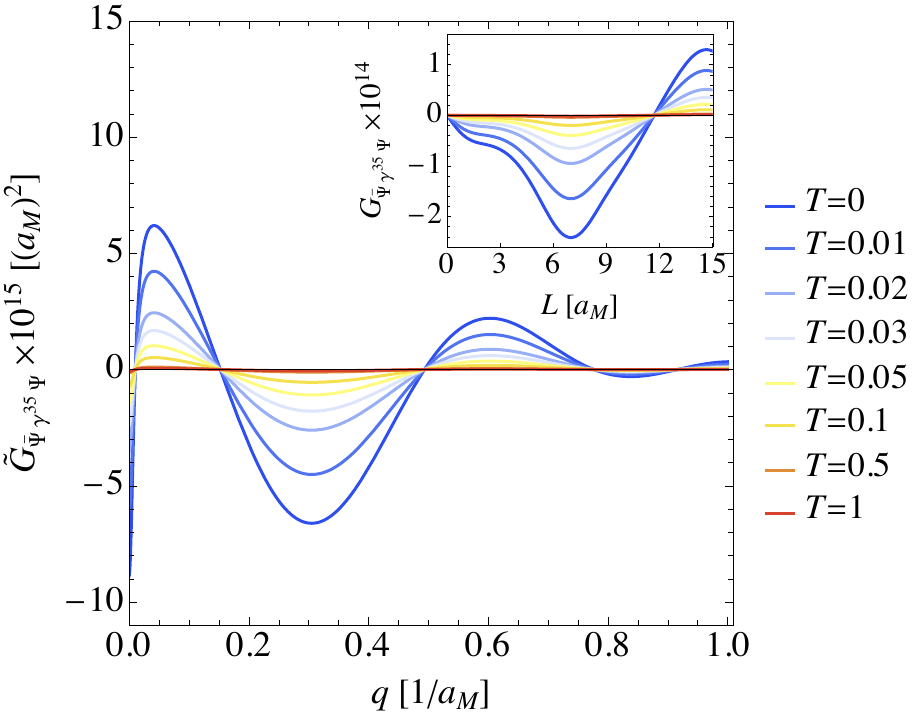}
\end{minipage}\qquad
\begin{minipage}[t]{.47\textwidth}
\includegraphics[height=0.77\linewidth]{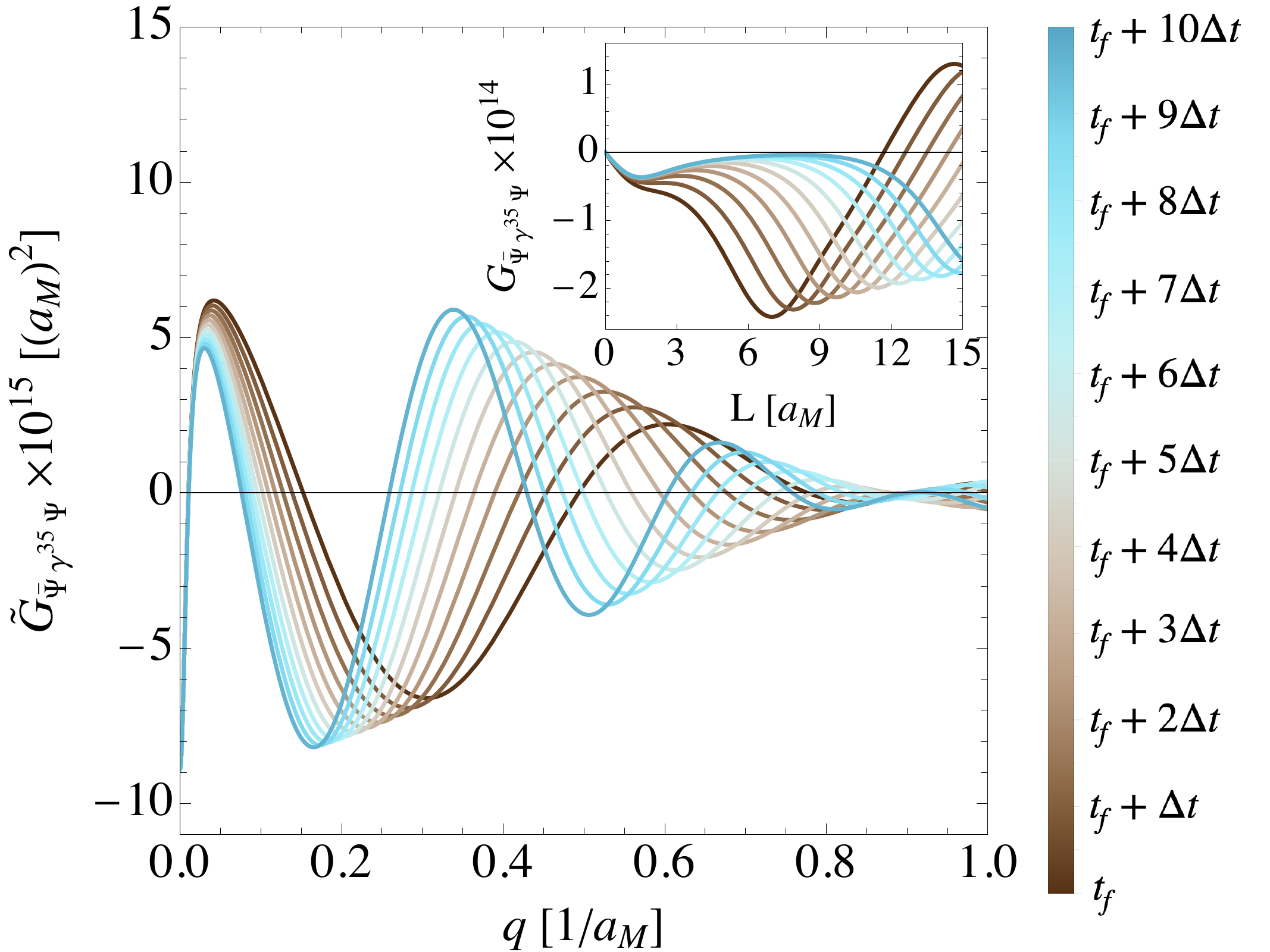}
\label{label-d}
\end{minipage}
\caption{\textbf{Equal-time two-point correlation function of fermionic excitations} in momentum space  $\tilde{\mathcal{G}}_{\bar{\Psi}\gamma^{35}\Psi}$ in units of $(a_M)^{2}$ as a function of the momentum $q$ in units of $1/a_M$. 
This has been calculated for a band gap term proportional to $\gamma^{35}$. 
The time dynamical process (region II) has a duration of $t_\text{f} - t_{\text{i}} = a_M/(5 v_0)$. The Fermi velocity $v_F$ is changed here from $v_0$ to $v_\text{f}=10^2 v_0$ and the band gap from $\Delta_0=50 \hbar v_0/a_M$ to $\Delta_\text{f}=\hbar v_0/a_M$ with a time dependence given in Eqs.~\eqref{eq:FermiVelocity} and \eqref{eq:Gap}. 
In the insets the dimensionless equal-time two-point correlation functions in position space as a function of the distance $L$  shown obtained after a regularization with a Gaussian window function of width $w=a_M$.
In the left panel, the colors correspond to different initial thermal states with temperature $T$ in units of $\hbar v_0/(k_\text{B} a_M)$.
The two-point correlation function is evaluated at final time $t_{\text{f}}$. 
In the right panel, the initial state is taken to be the vacuum at T=0K. The colors correspond to different holding times after the expansion has ceased with $\Delta t= a_M/(50 v_0)$.}
\label{fig:TwoPointCorrelationFunctionsForG35}
\end{figure*}

Two-point correlation functions are also a good indicator of particle production.
Let us start studying the equal-time two-point correlation functions in momentum space $\braket{\Bar{\Psi}_{\Vec{q}}\left(t\right)\gamma^\alpha\Psi_{\Vec{q^\prime}}\left(t\right)}$ for $\alpha=0,1,2$. 
The zero component of this set of two-point correlation functions corresponds to the electronic one-particle density matrix $\rho$, which evaluated at $t\geq t_\text{f}$ is given by
\begin{equation}
\begin{split}
     \rho \left(t;\Vec{q}\right) = 2,
\end{split}
\end{equation}
with $ \rho \left(t;\Vec{q}\right) (2\pi)^2\delta^{(2)}\left(\Vec{q}-\Vec{q^\prime}\right)=\braket{\Psi^\dagger_{\Vec{q}}(t) \Psi_{\Vec{q^\prime}}(t) }$
For the equal-time two-point correlation function with $\alpha=1$ evaluated at $t\geq t_\text{f}$
    \begin{equation}
\begin{split}
    \tilde{\mathcal{G}}_{\bar{\Psi} \gamma^{1}\Psi } (t,\Vec{q}) 
    =&\,  -2 
    \sum_\xi \xi \left( 1 - 2N_{\text{in}}^\xi \right)\\
    &\times \Big\{ \text{Re}\left[\tilde{u}^{\xi A*}_{\vec{q}} \tilde{u}^{\xi B}_{\vec{q}}\right] \left(2\abs{\beta^{\xi}_{\vec{q}}}^2 - 1 \right)\\
    &+ \text{Re}\left[ \left(\left(\tilde{u}^{\xi B}_{\vec{q}}\right)^2 - \left(\tilde{u}^{\xi A}_{\vec{q}}\right)^2\right) \alpha^{\xi}_{\vec{q}}\beta^{\xi}_{\vec{q}}\right] \Big\}\,,
  \label{eq:TwoPointCorrelationG1}
\end{split}
\end{equation}
with $\tilde{\mathcal{G}}_{\bar{\Psi} \gamma^{1}\Psi } (t,\Vec{q}) (2\pi)^2 \delta^{(2)}(\vec{q}-\vec{q'})  = i\braket{\bar{\Psi}_{\Vec{q}}(t) \gamma^{1} \Psi_{\Vec{q'}}(t) }$
and for the component $\alpha=2$
\begin{equation}
\begin{split}
     \tilde{\mathcal{G}}_{\bar{\Psi} \gamma^{2}\Psi } (t,\Vec{q}) 
     =&   -2
     \sum_\xi \left( 1 - 2N_{\text{in}}^\xi \right)\\
     &\times \Big\{ \text{Im}\left[\tilde{u}^{\xi A*}_{\vec{q}} \tilde{u}^{\xi B}_{\vec{q}}\right] \left( 2\abs{\beta^{\xi}_{\vec{q}}}^2 -1\right) \\
     &+ \text{Im}\left[\left(\left(\tilde{u}^{\xi B }_{\vec{q}}\right)^2 + \left(\tilde{u}^{\xi A }_{\vec{q}}\right)^2\right) \alpha^{\xi}_{\vec{q}}\beta^{\xi}_{\vec{q}}  \right] \Big\}\,,
\end{split}
\label{eq:TwoPointCorrelationG2}
\end{equation}
with $\tilde{\mathcal{G}}_{\bar{\Psi} \gamma^{2}\Psi } (t,\Vec{q}) (2\pi)^2 \delta^{(2)}(\vec{q}-\vec{q'})  = i\braket{\bar{\Psi}_{\Vec{q}}(t) \gamma^{2} \Psi_{\Vec{q'}}(t) }$. 
These two-point correlation functions in real space are related to the expectation value of the electric (vector) current evaluated at equal position, ${J_V^\alpha= -\Psi^\dagger \gamma^0 \gamma^{\alpha}\Psi}$~\cite{Parker1980,parker2009}, which vanish for $\alpha=1,2$ as a consequence of rotational symmetry.
Another two-point correlation function we consider is
\begin{equation}
\begin{split}
   \tilde{\mathcal{G}}_{\bar{\Psi} \gamma^{35}\Psi } (t,\Vec{q})  
   =&  - 
   \sum_\xi \xi \left( 1 - 2N_{\text{in}}^\xi \right) \\
     &\times \Big\{\left(2\abs{\tilde{u}^{\xi B}_{\vec{q}}}^2 - 1\right) \left(2\abs{\beta^{\xi}_{\vec{q}}}^2 - 1 \right)\\
     &- 4\text{Re}\left[ \tilde{u}^{\xi A}_{\vec{q}} \tilde{u}^{\xi B}_{\vec{q}} \alpha_{\vec{q}}^{\xi} \beta_{\vec{q}}^{\xi} \right] \Big\}\,,
 \label{eq:TwoPointCorrelationG35}
\end{split}
\end{equation}
with $  \tilde{\mathcal{G}}_{\bar{\Psi} \gamma^{35}\Psi } (t,\Vec{q}) (2\pi)^2 \delta^{(2)}(\vec{q}-\vec{q'}) = i\braket{\bar{\Psi}_{\Vec{q}}(t) \gamma^{35} \Psi_{\Vec{q'}}(t) }  $. We also study statistical two-point correlation functions at equal-time, e.g.,
\begin{equation}
\begin{split}
    \tilde{\mathcal{G}}^S_{\bar{\Psi} \Psi } (t,\vec{q})  =&  
    \sum_{\xi}  \left(1-2N^\xi_{\text{in}}\right)\\
    &\times \Big\{\left(2\abs{\tilde{u}^{\xi B}_{\vec{q}}}^2 - 1\right) \left( 2 \abs{\beta^{\xi}_{\vec{q}}}^2 - 1 \right) \\
     &- 4\text{Re}\left[\tilde{u}^{\xi A}_{\vec{q}} \tilde{u}^{\xi B}_{\vec{q}} \alpha_{\vec{q}}^{\xi} \beta_{\vec{q}}^{\xi} \right] \Big\}\,,
     \label{eq:StatisticalTwoPointCorrelationFunction}
\end{split}
\end{equation}
with $ \tilde{\mathcal{G}}^S_{\bar{\Psi} \Psi } (t,\vec{q}) (2\pi)^2 \delta^{(2)}(\vec{q}-\vec{q'}) = \frac{1}{2} \sum_{b} \braket{\left[\Bar{\Psi}^{b}_{\Vec{q}}(t),\Psi^{b}_{\Vec{q'}}(t) \right]}
  $, where the sum over $b$ indicates a sum over the components of the Dirac field.
For example, for the case of spinless fermions on a honeycomb lattice, the latter two-point correlation function provides information about the staggered density contributions on the sublattices. 

The two-point correlation functions related to superconductivity vanish for all points
\begin{equation}
\begin{split}
    &\braket{\Bar{\Psi}_{\Vec{q}}(t) \Psi^*_{\Vec{q'}}(t) } =0=\braket{\Bar{\Psi}^*_{\Vec{q}}(t) \Psi_{\Vec{q'}}(t) },
\end{split}
\end{equation}
and so do the ones corresponding to anomalous densities
\begin{equation}
\begin{split}
    &\braket{\Psi^\dagger_{\Vec{q}}(t) \Psi^*_{\Vec{q'}}(t)} =0=\braket{\Psi^T_{\Vec{q}}(t) \Psi_{\Vec{q'}}(t)}.
\end{split}
\end{equation}

Correlation functions in momentum space are regular, but an ultraviolet regularization is needed to represent them in position space (otherwise they would be distributions). To this end, we convolute the Dirac field in position space with a window function~\cite{Mukhanov2007}
\begin{equation}
    \Psi_W (t,\vec{x}) \equiv \int d^2x^\prime W(\vec{x}-\vec{x^\prime}) \Psi (t,\vec{x^\prime})\,.
\end{equation}
The window function is chosen as a normalized Gaussian 
\begin{equation}
    W (x) = \frac{1}{2\pi w^2} \exp{\left(-\frac{|\vec{x}|^2}{2w}\right)},
\end{equation}
with $w$ being the standard deviation or width which we choose to be $w=a_M$. 
In Fourier space, the window function acts as an ultraviolet regulator and it is also of Gaussian form, $\tilde{W}(q)=e^{-q^2w^2/2}$, such as the regularized expression for the two-point correlation functions becomes
\begin{equation}
\begin{split}
    \mathcal{G} (t,L) =& \int \frac{\text{d}^2q}{(2\pi)^2} e^{-i qL\cos\varphi} \tilde{\mathcal{G}} (t,\vec{q}) \tilde{W}^*(q) \tilde{W}(q),
\end{split}
\end{equation}
with $L=\abs{\vec{x}-\vec{x^\prime}}$ being the comoving distance between the two spatial positions $\vec{x}$ and $\vec{x^\prime}$. For $w\rightarrow0$ one formally recovers the full form of $\mathcal{G} (t,L)$ as a distribution.

In Fig.~\ref{fig:StatisticalTwoPointCorrelationFunction}, the equal-time two-point correlation function $\tilde{\mathcal{G}}_{\bar{\Psi} \gamma^{35}\Psi }$ and the statistical equal-time two-point correlation function $\tilde{\mathcal{G}}^S_{\bar{\Psi} \Psi }$ given in Eqs.~\eqref{eq:TwoPointCorrelationG35} and \eqref{eq:StatisticalTwoPointCorrelationFunction} respectively, are shown  for a band gap proportional to the identity after the time dynamical process. 
On the left panel, the initial state is taken as a thermal state for different temperatures and one can observe again the effect of Pauli blocking. 
On the right panel, the time evolution of the statistical equal-time two-point correlation function at $T=0$K is shown after the time dynamical process has ceased. 
Here, one can observe that the characteristic features of the two-point correlation function evolve with twice the speed of the final Fermi velocity $v_{\text{f}}$.

In Fig.~\ref{fig:TwoPointCorrelationFunctionsForG35}, the two-point correlation functions $\tilde{\mathcal{G}}_{\bar{\Psi} \gamma^{35}\Psi }$ is shown in momentum space and in real space for different initial thermal states (left panel) and for different holding times (right panel) after the time-dependent process has ceased for a band gap proportional to $\gamma^{35}$ given in Eq.~\eqref{eq:TwoPointCorrelationG35}. 
The characteristic features of this two-point correlation function evolve also with twice the final Fermi velocity. 
The evaluation of the statistical two-point correlation function, cf. Eq.~\eqref{eq:StatisticalTwoPointCorrelationFunction}, for an initial no-particle state vanishes for all distances for this gap shape.

The equal-time two-point correlation functions in momentum space $\tilde{\mathcal{G}}_{\bar{\Psi} \gamma^{1} \Psi }$ and $\tilde{\mathcal{G}}_{\bar{\Psi} \gamma^{2} \Psi }$ given in Eqs.~\eqref{eq:TwoPointCorrelationG1},~\eqref{eq:TwoPointCorrelationG2} depend on the direction of momentum in contrast to $\tilde{\mathcal{G}}_{\bar{\Psi} \gamma^{35}\Psi }$ and $\tilde{\mathcal{G}}^S_{\bar{\Psi} \Psi }$, which depend only on the momentum amplitude. 
Therefore, they have not been plotted in order to not choose a specific direction of the momentum.

\section{Conclusions and outlook}\label{sec:outlook}

In summary, we discussed how a situation analogous to cosmological particle production in an expanding spacetime could be realized in highly-tunable Dirac materials as they may be realized in moir\'e heterostructures.
The key ingredient is a controllable time-dependent ratio of an energy gap and the Fermi velocity. 
The energy gap, or mass term in relativistic nomenclature, is important because massless relativistic fermions are invariant under a scaling symmetry that would allow to absorb a time-dependent change in the metric scale factor or Fermi velocity.
In the presence of a finite gap this changes and a time dependence of the ratio $\Delta/v_F$ leads to the production of quasiparticle excitations.

We have investigated a scenario with unbroken charge conjugation symmetry and, here, the number of particle and antiparticle excitations are precisely equal for every momentum mode. 
At non-zero temperature, when some modes have a thermal occupation, particle production is suppressed by Pauli blocking.

Interesting observables to confirm and further investigate fermionic quasiparticle production are various two-point correlation functions as we exhibited in this contribution, which depend on the moiré Dirac material chosen as the analog platform. 
Their dependence on time as well as wavenumber contains characteristic information that could be compared between theory and a possible experiment.
It is of particular interest to reproduce these results experimentally to test the validity of highly tunable Dirac materials as quantum simulators. 

In the present paper we have not discussed the possibilities for experimental realizations of our proposal in much detail. 
However, we believe that moir\'e materials are particularly interesting in this regard.
A notorious example of moir\'e Dirac materials is twisted bilayer graphene, whose low energy fermionic excitations at charge neutrality have a Fermi velocity $v_{F}(\theta)$ which is a function of the twist angle $\theta$ originating from the interlayer coupling. Generically, the interlayer coupling in this van-der-Waals structure can be considered weak. However, it turns out that near the magic angle $\theta_M\sim 1.1^\circ$, the Fermi velocity $v_{F}(\theta)$ is strongly modified and approaches zero \cite{Bistritzer2011,LopesDosSantos2012,Cao2018}. The magic angle itself is determined by the interlayer coupling, which can be increased by applying pressure $\delta p$ to the system~\cite{Yankowitz2019,PhysRevB.98.085144}, i.e. $\theta_M=\theta_M(\delta p)$. Hence, even at a fixed angle near $\theta_M$, the Fermi velocity of the system can be strongly modified by an external perturbation, since it is very sensitive to any change in the system. 
In other words, around the magic angle, the Fermi velocity of TBG's Dirac excitations can be tuned over a wide range of values, potentially covering more than an order of magnitude. 
This may be experimentally achieved in a TBG system with a twist close to a magic angle~$\theta_M$ by different methods, e.g., (1) by varying the hydrostatic pressure $\delta p(t)$ with respect to time; (2) by applying ultrafast (non-interacting) light pulses, which produce the same mechanical effect as hydrostatic pressure; and (3) by performing continuous rotation, which has been achieved at room temperature \cite{Inbar2023}, but work is currently underway to achieve this at low temperatures. We think that the second one can be performed in the right time frame to bring the system out of equilibrium.
Another interesting material system could be the case of Bernal-stacked bilayer graphene, which hosts Dirac cones due to trigonal warping~\cite{mccann2013electronic}. Experimentally, the Dirac cones can be gapped out in a controlled way~\cite{seiler2022quantum}, which could possibly realize a scenario where the Fermi velocity is fixed and the gap is time-dependent. 
An alternative to choosing bilayer graphene to perform the experiment could be to design a honeycomb optical lattice to trap cold atoms. 
Time dependencies could be introduced into the system by modifying the potential depths in the sublattices, leading to a modification of the band gap. Therefore, in this case, all the time dependence would be contained in the mass term, but as demonstrated in the paper, we only need a time-dependent ratio between the mass and the Fermi velocity, so that would be enough to produce particles.

For future work, many extensions of the present setup are conceivable. One can add electromagnetic fields to study electromagnetic response and correlation functions in more detail. This would allow us to study the analog of Schwinger effect, for example. Further, it would be interesting to study also spatial curvature, which may be realized by making the Fermi velocity space dependent, or through inhomogeneous lattice configurations. Also, it would be highly interesting to investigate whether an analog to the production of the baryon-antibaryon asymmetry in the early universe could be realized in Dirac materials. In analogy to the famous Sakharov conditions this would likely need an explicit breaking of charge conjugation symmetry as well as time-reversal. Finally, it would be nice to study dynamically evolving spacetime geometries in Dirac materials and a possible interplay between the geometric and electronic degrees of freedom.

\section*{Acknowledgments}

The authors would like to thank Mar\'ia~A.~H. Vozmediano, Alberto Cortijo and Ilya Eremin for useful discussions.
MMS and MTS acknowledge funding from the Deutsche Forschungsgemeinschaft (DFG, German Research Foundation) within Project-ID 277146847, SFB 1238 (project C02). MMS is supported by the DFG Heisenberg programme (Project-ID 452976698).


\appendix

\section{Graphene representation and reducibility}\label{app:rep}

The set of time- and space-like gamma matrices ($\gamma^\alpha$ with $\alpha=0,1,2$) in the graphene representation are block diagonal, but the two blocks differ in $\gamma^1$. This shows that it is a reducible representation from the point of view of the Lorentz group and it is composed out of an irreducible representation with the two-component spinor $\begin{pmatrix}
    \psi^{+A}, &
    \psi^{+B}
\end{pmatrix}^T$ and a second irreducible representation that differs by a parity transform ${x^1\rightarrow -x^1}$. It is possible to perform a similarity transform in the Clifford algebra to make the reducibility manifest. Define the transformation matrix
\begin{equation}
    \hat{\gamma}_\pm = \begin{pmatrix}
    I_2 & 0\\
    0& \pm \sigma_2
    \end{pmatrix} \quad \text{with} \quad \left(\hat{\gamma}_\pm\right)^2 = I_4,
\end{equation}
and introduce the transformed representation of the Clifford algebra
\begin{equation}
    \Tilde{\gamma}^\mu = \hat{\gamma}_{-} \gamma^\mu \hat{\gamma}_{+},
\end{equation}
which is concretely $\tilde{\gamma}^0=I_2\otimes i\sigma^3$, $\tilde{\gamma}^1=I_2\otimes i\sigma^2$ and $\tilde{\gamma}^2=I_2\otimes (-\sigma^1)$. The transformed spinors can be taken as
\begin{equation}
\begin{split}
    &\Tilde{\Psi} = \hat{\gamma}_{+} \Psi,\\
    &\bar{\Tilde{\Psi}} = \bar{\Psi} \hat{\gamma}_{-} =i\Psi^\dagger \gamma^0 \hat{\gamma}_{-} = i \left( \hat{\gamma}_{+} \Psi \right)^\dagger \gamma^0.
\end{split}
\end{equation}
The Dirac-like action is invariant under the transformation. It becomes now clear that the transformed spinor
\begin{equation}
    \Tilde{\Psi} = \begin{pmatrix}
    \psi^{+A} \\
    \psi^{+B} \\
   -i \psi^{-B} \\
   +i \psi^{- A}
    \end{pmatrix}
\end{equation}
can be seen as two copies, or families, of Lorentz group irreducible two-component spinors. It becomes sensible to organize the spinors as 
\begin{equation}
    \Psi^\xi = \begin{pmatrix}
    \tilde{\psi}^{\xi A} \\
    \tilde{\psi}^{\xi B}
    \end{pmatrix},
    \label{eq:IrreducibleSpinor}
\end{equation}
where $\xi$ is the additional family index.

\section{Dirac field, Dirac matter metric and energy-momentum tensor}\label{app:General}

The Dirac field $\Psi$ can be quantized by introducing the equal-time anticommutation relations \cite{parker2009}
\begin{equation}
    \left\{\Psi_{a}(t,\vec{x}),\pi_{b}(t,\vec{x'})\right\} = i \hbar \delta^{(2)}(\vec{x}-\vec{x'}) \delta_{ab},
    \label{eq:AnticommutatorFieldMomentum}
\end{equation}
where $a, b$ refer to the spinor indices and the conjugate momentum density $\pi(t,\vec{x})$ is given by
\begin{equation}
    \pi(t,\vec{x}) = \frac{\partial \mathcal{L} [\Psi,\bar{\Psi}]}{\partial \left(\partial_t\Psi\right)} = -\sqrt{g} \hbar\bar{\Psi}(t,\vec{x}) \gamma^0 e_0^{\phantom{0}t}
    (t,\vec{x}).
\end{equation}
Therefore, the anticommutator \eqref{eq:AnticommutatorFieldMomentum} can be written as
\begin{equation}
    \left\{\Psi_{a}(t,\vec{x}),\bar{\Psi}_{b}(t,\vec{x'})\right\} =  -\frac{i}{\sqrt{g}}  \delta^{(2)}(\vec{x}-\vec{x'}) \left[\gamma^0 e_0^{\phantom{0}t}
    (t,\vec{x})\right]^{-1}_{ab}.
\label{eq:AnticommutatorDiracField}
\end{equation}

The energy-momentum tensor for a Dirac field is obtained by the variation of the action with respect to the spacetime metric \cite{Birrell1982}
\begin{equation}
\begin{split}
    \mathcal{T}^{\mu\nu} =& \frac{2}{\sqrt{g}} \frac{\delta S}{\delta g_{\mu\nu}}\, .
\end{split}
\end{equation}
For an action corresponding to the one studied here, cf. Eq.~\eqref{eq:vielbeinFieldsCartesian}, it reads
\begin{equation}
\begin{split}
    \mathcal{T}^{\mu\nu}=&  -\bar{\Psi}\hbar \gamma^\alpha e_\alpha^{\phantom{\alpha}\mu} \left(\partial^\nu + \Omega^\nu  \right)\Psi\\
    &+ g^{\mu\nu} \bar{\Psi}\left(\hbar \gamma^\alpha e_\alpha^{\phantom{\alpha}\sigma} \partial_\sigma + \Delta_j \Gamma^j \right) \Psi \, ,
\label{eq:EnergyMomentumTensor}
\end{split}
\end{equation}
with the diagonal components being (no sum implied)
\begin{equation}
\begin{split}
    \mathcal{T}^{\alpha}_{\phantom{\alpha}\alpha} &=  \bar{\Psi} \left( \hbar v_F \gamma^\alpha  \partial_\alpha + \Delta_j \Gamma^j \right) \Psi \, .
\end{split}
\end{equation}
In particular, the time-time component of the energy-momentum tensor leads to the Dirac Hamiltonian density, which in our case is given by
\begin{equation}
\begin{split}
    \mathcal{T}^{\eta}_{\phantom{\eta}\eta} \equiv \mathcal{H} =  \Bar{\Psi} \left(\hbar v_F \vec{\gamma}\cdot\vec{\partial} + \Delta_j \Gamma^j \right)\Psi \,.
\end{split}
\end{equation}

\section{Vielbein, metric and connections}\label{app:WeylTransformation}
We investigate here a vielbein field $e^\alpha_{\phantom{\alpha}\mu}(x)$ with the non-vanishing temporal component $e^0_{\phantom{0}0}(t) = e^{\zeta(t)} v_F(t)$, spatial component $e^1_{\phantom{1}1}(t) = e^2_{\phantom{2}2}(t) =e^{\zeta(t)}$, and all non-diagonal components vanishing. Here $v_F(t)$ is the Fermi velocity, and $e^{\zeta(t)}$ is a scaling factor factor that we keep unspecified for the time being. The corresponding spacetime metric $g_{\mu\nu}(x) = e^\alpha_{\phantom{\alpha}\mu}(x) e^\beta_{\phantom{\beta}\nu}(x) \eta_{\alpha\beta}$ has the non-vanishing components $g_{00}(t)= - e^{2\zeta(t)}v_F^2(t)$ and $g_{11}(t)=g_{22}(t) = e^{2\zeta(t)}$.

For this metric with $e^{\zeta(t)}=1$, i.e. Eq.~\eqref{eq:vielbeinFieldsCartesian}, one can determine the non-vanishing Christoffel symbol
\begin{equation}
\label{eq:ChristoffelSymbolsFlat}
    \Gamma^{t}_{\phantom{t}tt}= \frac{\dot{ v}_F(t)}{v_F(t)},
\end{equation}
with dot notation indicating a partial derivative respect to real time.
All spin connection components vanish.
Thus, this metric corresponds to a Minkowski spacetime with time-dependent Fermi velocity leading to a zero Ricci curvature.

For this metric with $e^{\zeta(t)}=\frac{\Delta_j(0) v_F(t)}{\Delta_j(t) v_F(0)}$, i.e. Eq.~\eqref{eq:GrapheneLineElement}, one can obtain the non-vanishing Christoffel symbols
\begin{equation}
\begin{split}
\label{eq:ChristoffelSymbolsFlat}
     \Gamma^{t}_{\phantom{t}xx}&=  \frac{1}{v_F^2(t)} \left(\frac{\dot{ v}_F(t)}{v_F(t)} - \frac{\dot{ \Delta}_j(t)}{\Delta_j(t)} \right) = \Gamma^{t}_{\phantom{t}yy} ,
    \\
     \Gamma^{x}_{\phantom{t}t x}&= \frac{\dot{ v}_F(t)}{v_F(t)}-  \frac{\dot{ \Delta}_j(t)}{\Delta_j(t)} =\Gamma^{x}_{\phantom{t}x t}=\Gamma^{y}_{\phantom{t} yt}=\Gamma^{y}_{\phantom{t}y t} ,\\
     \Gamma^{t}_{\phantom{t}t t}&= 2\frac{\dot{ v}_F(t)}{v_F(t)} -  \frac{\dot{ \Delta}_j(t)}{\Delta_j(t)}
\end{split}
\end{equation}
Therefore, the non-vanishing spin connection coefficients are
\begin{equation}
    \omega_{x01}= \frac{1}{v_F(t)} \left(\frac{\dot{ \Delta}_j(t)}{\Delta_j(t)} -\frac{\dot{ v}_F(t)}{v_F(t)} \right)  = \omega_{y02}  
\end{equation}
and their respective antisymmetric components.
Thus,the non-zero components of the spin connection are given by
\begin{equation}
\begin{split}
    &\Omega_{x} = 
\frac{1}{2 v_F(t)} \left(\frac{\dot{ \Delta}_j(t)}{\Delta_j(t)} -\frac{\dot{ v}_F(t)}{v_F(t)} \right) \gamma^0\gamma^1, \\
&\Omega_{y} = \frac{1}{2 v_F(t)} \left(\frac{\dot{ \Delta}_j(t)}{\Delta_j(t)} -\frac{\dot{ v}_F(t)}{v_F(t)} \right)  \gamma^0\gamma^2.
\end{split}
\end{equation}
The Ricci curvature, ${\mathcal{R}=g^{\mu\nu}R_{\mu\nu}}$, is given by
\begin{equation}
\begin{split}
    \mathcal{R}= 2\frac{v_F^2(0) \Delta_j^2(t)}{v_F^2(t) \Delta_j^2(0)} \Bigg[&- 3\frac{\dot{v}^2_F(t)}{v_F^2(t)} + \frac{\dot{\Delta}^2_j(t)}{\Delta^2_j(t)}\\
    &- 2\frac{\Ddot{\Delta}_j(t)}{\Delta_j(t)} + 2\frac{\Ddot{v}_F(t)}{v_F(t)}  \Bigg],
\end{split}
\end{equation}
which indicates a curvature deviation from an Euclidean spacetime when there is a band gap and a time dependence on the band gap and/or the Fermi velocity.

\section{General band gap}\label{app:GeneralBandGap}

\subsection{Mode equations and Hamiltonians}

Let us introduce the corresponding mode equations and Hamiltonians for different tensor shapes of the gap.

\subsubsection{Haldane mass}\label{sec:MassGamm35}

Mass terms proportional to $\gamma^{35}$ are block diagonal and, consequently, do not mix the Dirac points, such that one can study the mode equations for the different valleys independently, as we did previously with a mass term proportional to the identity matrix. 

Taking the mass gap tensor as $\Gamma^j=i\gamma^{35}$, the coupled system of first order differential equations~\eqref{eq:DiracEquationMass} for a Dirac field expanded in the basis~\eqref{eq:ExpansionDiracField} becomes
\begin{equation}
\begin{split}
0=\begin{pmatrix}
 i \partial_\eta - \xi \frac{ \Delta}{\hbar v_F} & \xi q e^{- \xi i\varphi}\\
  -\xi q e^{ \xi i\varphi}&  -i \partial_\eta - \xi \frac{ \Delta}{\hbar v_F}
\end{pmatrix}
\begin{pmatrix}
 u^{\xi A}_{\vec{q}}\\
 u^{\xi B}_{\vec{q}}
\end{pmatrix} \, .
\label{eq:DiracEquationModeFunctionsG35}
\end{split}
\end{equation}
Equivalently, the second order differential equation~\eqref{eq:SecondOrderRedefinedMassiveNonStaticDiracConformalGeneralGap} for this kind of mass term yields
\begin{equation}
\begin{split}
    &0= \left[\partial_\eta^2 + \omega^2_q(\eta) \varpm \xi
  \frac{i}{\hbar} \partial_\eta \left( \frac{ \Delta(\eta)}{ v_F(\eta)}\right) \right] u^{\xi A (B)}_{\vec{q}}(\eta) \, ,
    \label{eq:ConformalModeEquationsG35}
\end{split}
\end{equation}
which again corresponds to a harmonic oscillator differential equation with complex and time-dependent frequency.

For this kind of band gap, the massive part of the Hamiltonian, which corresponds to the Haldane mass in the case of spinless fermions on a honeycomb lattice, is given by
\begin{equation}
\begin{split}
    H_{\text{Hald.}} =& i \Delta \,\int \text{d}^2x \,\bar{\Psi}\,\gamma^{35} \Psi\\
    =& -\Delta \int \frac{\text{d}^2 q}{(2\pi)^2} \sum_\xi \xi \\
    &\times\big[ \left(\abs{u^{\xi B}_{\vec{q}}}^2  - \abs{u^{\xi A}_{\vec{q}}}^2 \right) \left( c^{\xi\dagger}_{\vec{q}} c^\xi_{\vec{q}} -  d^{\xi}_{\vec{q}} d^{\xi\dagger}_{\vec{q}} \right)\\ 
    & - 2 u^{\xi A }_{\vec{q}} u^{\xi B}_{\vec{q}} d_{\vec{-q}}^{\xi} c_{\vec{q}}^{\xi} - 2 u^{\xi A *}_{\vec{q}} u^{\xi B*}_{\vec{q}} c_{\vec{q}}^{\xi\dagger} d_{\vec{-q}}^{\xi\dagger} \big]\, ,
\end{split}
\end{equation}
where the Dirac field has been expanded again in the basis of Eq.~\eqref{eq:ExpansionDiracField}.
Thus, the Hamiltonian of the system reads
\begin{equation}
\begin{split}
   H=&  \int \frac{\text{d}^2 q}{(2\pi)^2}  \sum_\xi  \big[ E_{\vec{q}}^{\xi}(\eta) \left( c^{\xi \dagger}_{\vec{q}} c^{\xi }_{\vec{q}} - d^{\xi }_{\vec{q}} d^{\xi \dagger }_{\vec{q}}\right)\\
   &+ F^{\xi}_{\vec{q}}(\eta) d^{\xi }_{-\vec{q}} c^{\xi }_{\vec{q}} + F^{\xi*}_{\vec{q}}(\eta) c^{\xi \dagger}_{\vec{q}} d^{\xi \dagger}_{-\vec{q}}  \big]\, ,
\end{split}
\end{equation}
where now the functions $E^{\xi}_{\vec{q}}$ and $F^{\xi}_{\vec{q}}$ are given by
\begin{equation}
\begin{split}
    E_{\vec{q}}^{\xi} =& -2\xi \hbar v_F q \text{Re}\left[e^{-\xi i \varphi} u^{\xi A *}_{\vec{q}} u^{\xi B}_{\vec{q}}\right]\\
    &- \xi\Delta \left(1-2\left\lvert u^{\xi A}_{\vec{q}} \right\rvert^2\right),\\
    F_{\vec{q}}^{\xi} =& \xi\hbar v_F q \left[ e^{\xi i \varphi} \left(u^{\xi A}_{\vec{q}}\right)^2 - e^{-\xi i \varphi} \left(u^{\xi B}_{\vec{q}}\right)^2 \right] \\
    &+ 2 \xi \Delta u^{\xi A}_{\vec{q}} u^{\xi B}_{\vec{q}}\, .
    \label{eq:FunctionsHamiltonian35}
\end{split}
\end{equation}
The Bogoliubov coefficients are obtained by diagonalizing the Hamiltonian of the system through the Bogoliubov transformation \eqref{eq:BogoliuvobTransformationOperators} after the dynamical process has ceased. As this kind of mass gap is block diagonal, the Bogoliubov coefficients are also given by Eqs.~\eqref{eq:BogoliubovCoefficients} for the corresponding functions \eqref{eq:FunctionsHamiltonian35}.

A possible initial configuration of the mode functions leading to the no-particle initial state, i.e. an initial diagonal Hamiltonian, for this kind of gap term is given by
\begin{equation}
\begin{split}
   u^{\xi A, \text{I}}_{\vec{q}} (\eta\leq\eta_\text{i}) &= \frac{e^{-\xi i\varphi/2}}{\sqrt{2}}\sqrt{1 + \xi\frac{ \Delta_0}{\hbar v_0 \omega^{\text{I}}_q}} e^{-i\omega^{\text{I}}_q \eta} \, ,\\
    u^{\xi B, \text{I}}_{\vec{q}} (\eta\leq\eta_\text{i}) &=-\xi \frac{e^{\xi i\varphi/2}}{\sqrt{2}}\sqrt{1 - \xi \frac{ \Delta_0}{\hbar v_0 \omega^{\text{I}}_q}} e^{-i\omega^{\text{I}}_q \eta} \, .
    \label{eq:InitialConditionsG35}
\end{split}
\end{equation}
In region III, the set of mode functions associated to the creation and annihilation operators that diagonalize the Hamiltonian is given by
\begin{equation}
\begin{split}
   \tilde{u}^{\xi A, \text{III}}_{\vec{q}} (\eta\geq\eta_\text{f}) &= \frac{e^{-\xi i\varphi/2}}{\sqrt{2}}\sqrt{1 + \xi \frac{ \Delta_\text{f}}{\hbar v_\text{f} \omega^{\text{III}}_q}} e^{-i\omega^{\text{III}}_q \eta} \, ,\\
    \tilde{u}^{\xi B, \text{III}}_{\vec{q}} (\eta\geq\eta_\text{f}) &= -\xi \frac{e^{\xi i\varphi/2}}{\sqrt{2}}\sqrt{1 - \xi \frac{ \Delta_\text{f}}{\hbar v_\text{f} \omega^{\text{III}}_q}} e^{-i\omega^{\text{III}}_q \eta} \, .
    \label{eq:FinalConditionsG35}
\end{split}
\end{equation}

\subsubsection{Kekul\'e modulation of the nearest neighbor hopping}\label{sec:MassKekule}
    
A mass term $ i\Delta(\gamma^3\cos\theta + \gamma^5\sin\theta)$, with real amplitude $\Delta$ and phase $\theta$, mixes the Dirac points and the sublattices in the case of spinless fermions on a honeycomb lattice. Consequently, the mode functions for the different Dirac points cannot be studied independently as in the previous Secs.~\ref{sec:ModeEquations} and \ref{sec:MassGamm35}.

For a band gap proportional to $i(\gamma^3\cos\theta + \gamma^5\sin\theta)$ and taking into consideration the expansion of the Dirac field in Fourier modes \eqref{eq:ExpansionDiracField},   the mode equation is 
\begin{equation}
\begin{split}
0=\begin{pmatrix}
 i \partial_\eta  & q e^{-i\varphi} & 0 &   \frac{\bar{\Delta}^*}{\hbar v_F}\\
  -q e^{i\varphi} & - i \partial_\eta  &  - \frac{\bar{\Delta}^*}{\hbar v_F} & 0\\
   0 &   \frac{\bar{\Delta}}{\hbar v_F} & i \partial_\eta  &  - q e^{i\varphi}  \\
    -\frac{\bar{\Delta}}{\hbar v_F} & 0 &  q e^{-i\varphi}  & -i \partial_\eta 
\end{pmatrix}
\begin{pmatrix}
 \vphantom{\partial_\eta}u^{+A}_{\vec{q}}\\
\vphantom{\partial_\eta}    u^{+B}_{\vec{q}}\\
\vphantom{\partial_\eta}    u^{- A}_{\vec{q}}\\
\vphantom{\partial_\eta}    u^{- B}_{\vec{q}}
\end{pmatrix} \, ,
\end{split}
\end{equation}
with $\bar{\Delta}=\Delta e^{i\theta}$ and, equivalently, Eq.~\eqref{eq:SecondOrderRedefinedMassiveNonStaticDiracConformalGeneralGap}
\begin{equation}
\begin{split}
    &0=\left[\partial_\eta^2  + \omega^2_q(\eta)   \right] u^{+ \lambda}_{\vec{q}}(\eta) + \frac{i}{\hbar} \partial_\eta \left(\frac{ \bar{\Delta}^*(\eta)}{ v_F(\eta)} \right) u^{- \lambda^\prime}_{\vec{q}}(\eta)\, ,\\
    &0=\left[\partial_\eta^2   + \omega^2_q(\eta)   \right] u^{-\lambda}_{\vec{q}}(\eta) + \frac{i}{\hbar} \partial_\eta \left(\frac{ \bar{\Delta}(\eta)}{ v_F(\eta)} \right) u^{+ \lambda^\prime}_{\vec{q}}(\eta)\, ,
\end{split}
\end{equation}
where the superindex $\lambda\neq\lambda^\prime$.

The corresponding massive part of the Hamiltonian is given
\begin{equation}
\begin{split}
     H_{\text{Kekul\'e}} =& i\int \text{d}^2x \,\Psi^\dagger \begin{pmatrix}
         0 & \bar{\Delta}^* \sigma^1\\
         \bar{\Delta} \sigma^1 & 0
     \end{pmatrix} \Psi.\\
\end{split}
\end{equation}
Therefore, the Hamiltonian of the system
\begin{equation}
\begin{split}
   H 
   =& \int \frac{\text{d}^2 q}{(2\pi)^2} \Bigg\{ \sum_\xi  \Big[ E_{\vec{q}}^{(0)\xi} \left( c^{\xi \dagger}_{\vec{q}} c^{\xi }_{\vec{q}} - d^{\xi }_{\vec{q}} d^{\xi \dagger }_{\vec{q}}\right)\\
   &+ F_{\vec{q}}^{(0)\xi} d^{\xi }_{-\vec{q}} c^{\xi }_{\vec{q}} + F_{\vec{q}}^{(0)\xi*} c^{\xi \dagger}_{\vec{q}} d^{\xi \dagger}_{-\vec{q}}  \Big] \\
    & +  \Big[J_{\vec{q}} \left(c_{\vec{q}}^{-\dagger} c_{\vec{q}}^{+} - d_{\vec{-q}}^{+} d_{\vec{-q}}^{- \dagger} \right)\\
    &\hspace{0.4cm}+ G_{\vec{q}} \left(d_{\vec{-q}}^{-} c_{\vec{q}}^{+} + d_{\vec{-q}}^{+} c_{\vec{q}}^{-} \right) + \text{h.c.} \Big]\Bigg\}\, ,
    \label{eq:Hamiltonian3}
\end{split}
\end{equation}
with
\begin{equation}
\begin{split}
     &E_{q}^{(0)\xi} = -2\xi \hbar v_F q \text{Re}\left[e^{-\xi i \varphi} u^{\xi A *}_{\vec{q}} u^{\xi B}_{\vec{q}}\right] \, ,\\
    &F_{\vec{q}}^{(0)\xi} = \xi\hbar v_F q \left[ e^{\xi i \varphi} \left(u^{\xi A}_{\vec{q}}\right)^2 - e^{-\xi i \varphi} \left(u^{\xi B}_{\vec{q}}\right)^2 \right]\, ,\\
    &J_{\vec{q}} = i\bar{\Delta} \left(u_{\vec{q}}^{-A*}u_{\vec{q}}^{+B}  + u_{\vec{q}}^{-B*}u_{\vec{q}}^{+A} \right)\, ,\\
    & G_{\vec{q}} = i\bar{\Delta}  \left(u_{\vec{q}}^{+B}u_{\vec{q}}^{-B} - u_{\vec{q}}^{+A}u_{\vec{q}}^{-A} \right) \, ,
\end{split}
\end{equation}
is diagonalized by using the general Bogoliubov transformations given in Eq.~\eqref{eq:BogoliubovTransformationOperators}.

\section{Dirac spinor and Lorentz transformations}\label{app:lorentz}

The generators of the Lorentz group are identified as
\begin{equation}
    S^{\mu\nu} = \frac{i}{4} \left[\gamma^\mu,\gamma^\nu\right]\,.
\end{equation}
In a (2+1) dimensional spacetime this can be decomposed into a single spatial-spatial generator corresponding to a unique rotation 
\begin{equation}
    S^{12} = \frac{i}{2} \gamma^1 \gamma^2\,,
\end{equation}
and into two spatial-temporal generators corresponding to boosts
\begin{equation}
    S^{01} = \frac{i}{2} \gamma^0 \gamma^1 \quad \text{and} \quad S^{02} = \frac{i}{2} \gamma^0 \gamma^2\,.
\end{equation}
In the graphene representation ${S^{12} = - \sigma^3 \otimes \sigma^3/2}$ is diagonal, while ${S^{01} = i\sigma^3 \otimes \sigma^1/2}$ and ${S^{02} = i\sigma^0 \otimes \sigma^2/2}$.

A Dirac spinor transforms under Lorentz transformations as
\begin{equation}
    \Psi(x) \rightarrow \Psi^\prime(x^\prime)=S[\Lambda] \Psi (\Lambda^{-1}x)
    \label{eq:LorentzTransformationSpinor}
\end{equation}
and
\begin{equation}
    \bar{\Psi}(x) \rightarrow \bar{\Psi}^\prime(x^\prime)= \bar{\Psi} (\Lambda^{-1}x) S^{-1}[\Lambda],
\end{equation}
with $S[\Lambda]$ being the exponentiation of the Lorentz generators
\begin{equation}
\begin{split}
    S[\Lambda] = \exp\left(\frac{i}{2} \Omega_{\mu\nu} S^{\mu\nu}\right),
\end{split}
\end{equation}
where $\Omega_{\mu\nu}$ are six antisymmetric numbers which define the angle of rotation and the rapidity for the boosts.

\section{Parameter values}
\label{app:Values}

The plots shown in Sec.\ \ref{sec:Results} are computed for the following experimental parameters.
The moir\'e lattice constant is given by ${a_{\text{M}} = a_0/\left[2\sin(\theta/2)\right]}$, where $a_0$ is taken as the graphene lattice constant ${a_0\simeq2.46\angstrom}$ and for the reported narrow bands the twist angle is taken to be ${\theta\simeq1.1\degree}$, which leads to the moir\'e lattice constant ${a_{\text{M}} \simeq a_0/\theta \simeq 15\,\mathrm{nm}}$~\cite{Cea2022}. 
The Fermi velocity is changed from ${v_0=c/10^5}$ to ${v_\text{f}=c/10^3}$, where $c$ denotes the speed of light, while the mass gap from ${\Delta_0=50 \hbar v_0/a_M\simeq 16.9\,\mathrm{meV}}$ to ${\Delta_\text{f}=\hbar v_0/a_M\simeq 0.3\,\mathrm{meV}}$. 
The time dynamical process has a duration of ${t_\text{f}-t_\text{i}=a_M/(5v_0)= 10^{-12}\,\mathrm{s}}$ and the holding time steps ${\Delta t = a_M/(50v_0) = 10^{-13}\,\mathrm{s}}$. 
The initial temperatures $T$ are taken from $0\,\mathrm{K}$ to ${\hbar v_0/(k_\text{B}a_M)\simeq1.5\,\mathrm{K}}$. 
The window function for the regularization of the correlation functions in position space has Gaussian form with a width $w=a_M$.

\bibliography{references.bib}

\end{document}